\documentclass[preprint,12pt]{./elsarticle}

\usepackage{lineno,hyperref}

\usepackage{graphicx}
\usepackage{subcaption}
\usepackage{amssymb}
\usepackage{amsfonts}
\usepackage{amsmath} 
\usepackage{amsbsy} 
\usepackage{amsthm} 
\usepackage{mathrsfs} 
\usepackage{epstopdf} 
\usepackage{dsfont} 

\newtheorem{Theorem}{Theorem}
\newtheorem{hypothesis}{Hypothesis}
\newtheorem{corollary}{Corollary}
\newtheorem{remark}{Remark}
\newtheorem{condition}{Condition}

\begin{document}

\begin{frontmatter}

\title{A Shell Frictionally Coupled to an Elastic Foundation and a Comparison Against the Two-Body Coulomb's Law of Static Friction\tnoteref{t1}}
\tnotetext[t1]{This work is based on a PhD thesis submitted to UCL in 2016 \citep{jayawardana2016mathematical}, and the initial research was funded by The Dunhill Medical Trust [grant number R204/0511] and UCL Impact Studentship.}

\author[mymainaddress]{Kavinda Jayawardana\corref{mycorrespondingauthor}}
\address[mymainaddress]{TEK Optima Research Ltd, Unit 10 Westcroft Buiness Park Oakdene Drive, Three Legged Cross, Wimborne BH21 6FQ}
\ead{kavjayawardana@tekoptimaresearch.com; zcahe58@ucl.ac.uk}
\cortext[mycorrespondingauthor]{Corresponding author}

\begin{abstract}
In this article, we derive a model for a shell that is frictionally coupled to an elastic foundation. We use Kikuchi and Oden's model for Coulomb's law of static friction \cite{Kikuchi} to derive a displacement-based static-friction law for a shell on an elastic foundation model, and we prove the existence and the uniqueness of solutions with the aid of the work of  Kinderlehrer
and Stampacchia \cite{kinderlehrer2000introduction}. For numerical analysis, we modify Kikuchi and Oden's model for Coulomb's law of static friction \cite{Kikuchi} to model a full two-body contact problem in curvilinear coordinates. Our numerical results indicate that if the shell has a relatively high Young's modulus or has a relatively high Poisson's ratio, and the contact region has a high coefficient of friction or has a high radius of curvature, then the displacement field of the foundation predicted by both models are in better agreement. As far as we are aware, this is the first derivation of a displacement-based friction law and a two-body 3D elasticity contact problem with friction in the literature.
\end{abstract}


\end{frontmatter}


\section{Introduction}

Consider a situation where two elastic bodies are in contact with each other and the contact region exhibits friction, where friction defined as the force that oppose potential relative motion between the two bodies. A common area where modelling of such problems can be found is in the field of tyre manufacturing \citep{bowden2001friction, clark1981mechanics}. Assume now where one of the elastic bodies is very thin and almost planar in a curvilinear sense relative to the other body. Then the thin body can be approximated by a shell or a membrane, and such models can be used to model skin abrasion caused by fabrics as a result of friction \cite{jayawardana2017quantifying}. There is a need for valid modelling techniques in fields such as sports related skin trauma \citep{bergfeld1985trauma} and cosmetics \citep{asserin2000measurement}. It is documented that abrasion damage to human skin in cases such as the jogger's nipple \citep{levit1977jogger} and dermatitis from clothing \citep{wilkinson1985dermatitis} are caused by repetitive movement of fabrics on skin, and in cases such as pressure ulcers \citep{maklebust2001pressure} and juvenile plantar dermatitis \citep{shrank1979aetiology}, friction may worsen the problem. The aim of this article is to present a simple but a mathematically valid method to model such problems, i.e. a mathematical model for a shell on an elastic foundation when subjected to a displacement-based friction condition in a \emph{static dry-friction} (see section 11.3 of Kikuchi and Oden \cite{Kikuchi}) setting.

\subsection{Modelling Difficulties}

Consider a three-dimensional elastic body that is in contact with a rigid boundary whose contact area is \emph{rough}, i.e. contact area exhibits friction (see chapter 13 of Johnson \cite{johnson1987contact} or  section 5.2 of Quadling and Neill \cite{quadling2004mechanics}), then, given that we know the pressure experienced on the elastic body at the contact region in advance, the governing equations that describe the behaviour at the contact region can be represented by Kikuchi and Oden's model for Coulomb's law of static friction \cite{Kikuchi}, which has the following formulation in Euclidean coordinates
\begin{equation}\label{odenJepsi}
\qquad j_\varepsilon (\textbf{u}) = \left\{\begin{aligned}
\int_\Gamma \left[\mathscr{K}|\textbf{u}_T| - \frac{1}{2}\varepsilon\right] ds ,~& \text{if}~|\textbf{u}_T| \geq \varepsilon,\\
\int_\Gamma \left[\frac{1}{2} \mathscr{K} \frac{| \textbf{u}_T|^2}{\varepsilon} \right]ds ,~& \text{if}~|\textbf{u}_T| < \varepsilon,
\end{aligned}
\right.
\end{equation}
where  $\nu_F$ is the coefficient of friction, $\textbf{u}$ is the displacement field and $\textbf{u}_T$ is the tangential displacement field of the contact boundary, $\mathscr{K}$ (units: $\text{Nm}^{-2}$) is the spring modulus, $\varepsilon$ is the regularisation parameter, and $\Gamma$ is the contact region between elastic body and rigid obstacle. Let $\boldsymbol \sigma_T (\textbf u)$ and $ \sigma_n (\textbf u)$ be the normal-tangential stress and purely-normal stress tensors at the contact boundary respectively. Thus, if one assumes that the purely-normal stress, $\sigma_n(\textbf{u})<0$ (i.e. pressure), is no longer an unknown, but it is prescribed, and further assumes that $\mathscr{K} = - \nu_F \sigma_n(\textbf{u})$, then the G\^{a}teaux derivative (see definition 1.3.7 of Badiale and Serra \cite{badiale2010semilinear}) of $j_\varepsilon(\cdot)$ has the following form
\begin{equation} \label{CoulombOden}
\qquad \boldsymbol  \sigma _T (\textbf{u}) = \left\{\begin{aligned}
\nu_F \sigma_n(\textbf{u}) \frac{\textbf{u}_T}{|\textbf{u}_T|} , &~\text{if}~|\textbf{u}_T| \geq \varepsilon,\\
\nu_F \sigma_n(\textbf{u}) \frac{\textbf{u}_T}{\varepsilon} ,  &~\text{if}~|\textbf{u}_T| < \varepsilon.
\end{aligned}
\right.
\end{equation}
Now, assume that we are considering a shell (i.e. two-dimensional representation of a very thin three-dimensional elastic body), then the very idea of normal stress becomes meaningless. This is because for a shell, we find that  $\sigma _T (\textbf{u}) = \boldsymbol 0$ and $ \sigma_n(\textbf{u}) = 0$, and thus, Kikuchi and Oden's model \cite{Kikuchi}, i.e. equation (\ref{CoulombOden}), will fail to be applicable. Note that for a thorough mathematical analysis of the shell theory, consult chapter 4 of Ciarlet \cite{ciarlet2005introduction}.\\ 

One could find other 1D and 2D-elasticity models that incorporates friction, such as the capstan equation, belt-friction models \citep{rao2003engineering}, and beams with friction \citep{gao2000finite}; however, all such models (including Kikuchi and Oden's model \cite{Kikuchi}) deal with a rigid obstacle as the contact surface, and thus, at the presence of an elastic obstacle (i.e. a two-body elasticity contact problem with friction), all such friction models fail. Currently, there exist two-body 1D-elasticity contact models with friction in the literature (i.e. analysis of elastic strings in contact \cite{doonmez2004model,grandgeorge2021mechanics,maddocks1987ropes,warren2018clothes}). However, the literature still lacks two-body 2D and 3D-elasticity contact models with friction.

\section{A Shell Frictionally Coupled to an Elastic Foundation}

\begin{figure*}[!h]
\centering
\includegraphics[width=1\textwidth]{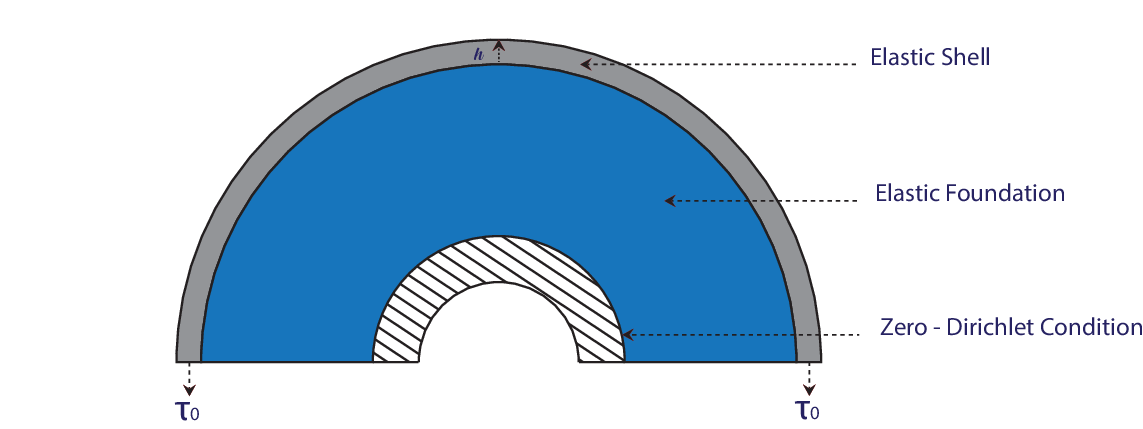}
\caption{Schematic representation of a shell frictionally coupled to an elastic foundation in Euclidean space, where $h$ is the thickness and $\tau_{\!0}$ is some traction acting on the boundary of the shell}
\label{shellonfoundation}
\end{figure*}

In this section, we modify  Kikuchi and Oden's model for Coulomb's law of static friction \cite{Kikuchi} to derive a friction condition to model the behaviour of an elastic shell that is frictionally coupled to an elastic-foundation (see fig. \ref{shellonfoundation}). To do so, consider an unstrained static three-dimensional isotropic elastic body (which we call the \emph{foundation}) whose volume is described by the diffeomorphism $\bar{\boldsymbol X} \in C^2(\bar\Omega;$ $\textbf{E}^3)$, where $\Omega\subset \mathbb{R}^3$ is a connected open bounded domain that satisfies the segment condition with a uniform-$C^1(\mathbb R^3;\mathbb R^2)$ boundary (see definition 4.10 of Adams and Fournier \cite{adams2003sobolev}), $C^k(\cdot)$ is a space of continuous functions that has continuous first $k$ partial derivatives in the underlying domain, $\textbf{E}^k$ is the $k$-dimensional Euclidean space and $\mathbb{R}^k$ is the $k$-dimensional curvilinear space. Now, assume that there exists a thinner isotropic elastic body (which we call the \emph{shell}, i.e. a planar elastic body with constant thickness $h$ whose rest configuration is curvilinear) frictionally coupled to a subset of the boundary of the foundation such that the contact region is initially stress-free, where this contact region is described by the injection $\boldsymbol \sigma \in C^3(\bar\omega;\textbf{E}^2)$ and where $\omega\subset \mathbb{R}^2$ is a connected open bounded plane that satisfies the segment condition with a uniform-$C^1(\mathbb{R}^2;\mathbb{R})$ boundary. Note that in our analysis, we only consider shells that satisfy the following condition:
\begin{condition}\label{dfnShell}
Let the map $\boldsymbol \sigma \in C^3(\bar\omega;\textbf{E}^2)$ describes the lower-surface of an unstrained shell, where $\omega \subset\mathbb{R}^2$ is a connected open bounded plane that satisfies the segment condition with a uniform-$C^1(\mathbb{R}^2;\mathbb{R})$ boundary. Given that the thickness of the shell is $h$, we require the following condition to be satisfied
\begin{align*}
0\leq h^2K<  hH\ll  1 , ~\forall ~ \boldsymbol(x^1,x^2\boldsymbol) \in \bar\omega,
\end{align*}
where $K$ is the Gaussian curvature and $H$ is the mean curvature, i.e. the lower-surface of the shell is non-hyperbolic and it is a surface with a positive mean curvature, and the thickness of the shell is sufficiently small.
\end{condition}
Note that $K = (F_{\!\text{[II]}1}^{~~1}F_{\!\text{[II]}2}^{~~2}-F_{\!\text{[II]}1}^{~~2}F_{\!\text{[II]}2}^{~~1})$, $H=-\frac12F_{\!\text{[II]}\alpha}^{~~\alpha}$,
\begin{align*}
F_{\!\text{[II]}\alpha\beta} =\boldsymbol  N  \cdot  \partial_{\alpha\beta}\boldsymbol \sigma, ~  ~\alpha,\beta\in \{1,2\},
\end{align*}
is the second fundamental form tensor of $\boldsymbol \sigma$, 
\begin{align*}
\boldsymbol N = \frac{\partial_1\boldsymbol \sigma \times \partial_2\boldsymbol \sigma}{||\partial_1\boldsymbol \sigma \times \partial_2\boldsymbol \sigma||}
\end{align*}
is the unit normal to the surface $\boldsymbol \sigma$, $\partial_\alpha$ are partial derivatives with respect to curvilinear coordinates $x^\alpha$, and $\cdot$, $\times$ and $||\cdot||$ are the Euclidean dot product, the Euclidean cross product and the Euclidean norm respectively. Also, note that Einstein's summation notation (see section 1.2 of Kay \cite{kay1988schaum}) is assumed throughout, bold symbols signify that we are dealing with vector and tensor fields, we regard the indices $i,j,k,l \in \{1,2,3\}$ and $\alpha,\beta,\gamma,\delta \in \{1,2\}$, and we usually reserve the vector brackets $\boldsymbol (\cdot \boldsymbol )_{\text{E}}$ for vectors in the Euclidean space and $\boldsymbol (\cdot \boldsymbol )$ for vectors in the curvilinear space.\\

With condition \ref{dfnShell} and in accordance with the work of Jayawardana \cite{jayawardana2016mathematical}, we can express the energy functional of a shell bonded to an elastic foundation as follows
\begin{align*}
J(\boldsymbol u) = &\int_{\Omega} \left[\frac{1}{2}A^{ijkl}E_{ij}(\boldsymbol u) E_{kl}(\boldsymbol u) - f^i u_i \right]d\Omega \\
&+\int_{\omega}\bigg[\frac{1}{2}B^{\alpha\beta\gamma\delta}\bigg(h\epsilon_{\alpha\beta}(\boldsymbol u) \epsilon_{\gamma\delta}(\boldsymbol u) + \frac{1}{3}h^3\rho_{\alpha\beta}(\boldsymbol u) \rho_{\gamma\delta} (\boldsymbol u) \bigg) - h f_0^i u_i \bigg]d\omega \\
&- \int_{\partial\omega} h \tau_{\!0}^i u_i ~d(\partial\omega) ,
\end{align*}
where $\boldsymbol u$ is the displacement field, $T^{ij}(\boldsymbol u) = A^{ijkl}E_{kl}(\boldsymbol u) $ is second Piola-Kirchhoff stress tensor  of the foundation, $E_{ij} (\boldsymbol u) = \frac{1}{2} (\bar\nabla_{\!i} u_j +\bar\nabla_{\!j} u_j)$ is  linearised Green-St Venant strain tensor  of the foundation, $$A^{ijkl} = \bar\lambda g^{ij} g^{kl} + \bar\mu (g^{ik}g^{jl} + g^{il}g^{jk})$$ is the isotropic elasticity tensor  of the foundation,
\begin{align*}
g_{ij} = \partial_i \bar{\boldsymbol{X}} \cdot \partial_j \bar{\boldsymbol{X}},  ~i,j \in \{1,2,3\},
\end{align*}
is the covariant metric tensor of $\bar{\boldsymbol X}$,  $\partial_j$ is the partial derivative with respect to the coordinate $x^j$, $$\bar\lambda = \frac{ \bar\nu \bar E }{(1+\bar\nu)(1-2\bar\nu)}$$ and $$\bar\mu = \frac{1}{2}\frac{\bar E}{1+\bar\nu}$$ are the first and the second Lam\'{e}'s parameters  of the foundation respectively, $\bar E \in(0,\infty)$ is the Young's modulus  of the foundation and $\bar\nu \in(-1,\frac12)$ is the Poisson's ratio of the elastic foundation, $\boldsymbol f$ is an external force density field acting on the elastic foundation, and $\bar{\boldsymbol n}$ is the unit outward normal to the boundary $\partial\Omega$ in curvilinear coordinates. Furthermore, $\tau^{\alpha\beta}(\boldsymbol u) = B^{\alpha \beta \gamma \delta} \epsilon_{\gamma \delta}(\boldsymbol u)$ is the stress tensor, $\eta^{\alpha\beta} (\boldsymbol u) = B^{\alpha \beta \gamma \delta} \rho_{\gamma \delta}(\boldsymbol u)$ negative of the change in moments density tensor of the shell, 
\begin{align*}
\epsilon_{\alpha\beta}(\boldsymbol u) = \left[\frac{1}{2} \left(\nabla_{\!\alpha} u_\beta+\nabla_{\!\beta} u_\alpha\right) - g_3 F_{\!\text{[II]}\alpha \beta} u^3\right]\!|_{\omega}
\end{align*}
is half of the change in the first fundamental form tensor of the shell and $g_3 = \sqrt{g_{33}}|_{\omega} = 1$ by construction, 
\begin{align*}
\rho_{\alpha\beta}(\boldsymbol u) = \Big[g_3\left(\nabla_{\!\alpha} \nabla_{\!\beta} u^3  - F_{\!\text{[II]}\alpha \gamma} F_{\!\text{[II]}\beta}^{~~\gamma} u^3\right) +F_{\!\text{[II]}\beta\gamma}\nabla_{\!\alpha} u^\gamma & \\
+ F_{\!\text{[II]}\alpha\gamma} \nabla_{\!\beta} u^\gamma +\left(\nabla_{\!\alpha}F_{\!\text{[II]}\beta \gamma }\right) u^\gamma & \Big] |_{\omega}
\end{align*}
the change in the second fundamental form tensor of the shell,
\begin{align*}
B^{\alpha \beta \gamma \delta} & = \frac{2\lambda\mu}{\lambda + 2\mu}F_{\!\text{[I]}}^{\alpha\beta}F_{\!\text{[I]}}^{\gamma \delta}+ \mu (F_{\!\text{[I]}}^{\alpha \gamma}F_{\!\text{[I]}}^{\beta\delta} + F_{\!\text{[I]}}^{\alpha\delta}F_{\!\text{[I]}}^{\beta\gamma})
\end{align*}
is the isotropic elasticity tensor of the shell,  $$\lambda = \frac{ \nu E }{(1+\nu)(1-2\nu)} $$ and $$\mu = \frac{1}{2}\frac{ E}{1+\nu}$$ are the first and the second Lam\'{e}'s parameters of the shell respectively, $E\in(0,\infty)$ is the Young's modulus of the shell and $\nu \in(-1,\frac12)$ is the Poisson's ratio of the shell,  $\boldsymbol f_0$ is an external force density field acting on the shell, $\boldsymbol n$ is the unit outward normal vector to the boundary $\partial\omega$ in curvilinear coordinates, $\boldsymbol \tau_{\!0}$ is an external traction field acting on the boundary of the shell, and $\boldsymbol u|_{\omega}$ is in a trace sense (see section 5.5 of Evans \cite{Evans}). Finally, $\bar{\boldsymbol \nabla}$ is the covariant derivative operator in the curvilinear space, i.e. for any $\boldsymbol v \in C^1(\bar\Omega;\mathbb{R}^3)$, we define its covariant derivative as follows
\begin{align*}
\bar \nabla_{\!j} v^k = \partial_j v^k + \bar \Gamma^k_{\!ij}v^i,
\end{align*}
where
\begin{align*}
\bar \Gamma^k_{\!ij} = \frac{1}{2}g^{kl}\left ( - \partial_l g_{ij} +\partial_i g_{jl} +\partial_jg_{li} \right)
\end{align*}
are the Christoffel symbols of the second kind, and $\boldsymbol \nabla$ is the covariant derivative operator in the curvilinear plane, i.e. for any $\boldsymbol u \in  C^1(\bar\omega;\mathbb{R}^2)$, we define its covariant derivative as follows
\begin{align*}
\nabla_{\!\beta} u^\gamma = \partial_\beta u^\gamma + \Gamma^\gamma_{\!\alpha\beta}u^\alpha,
\end{align*}
where
\begin{align*}
\Gamma^\gamma_{\!\alpha\beta} = \frac{1}{2}F_{\!\text{[I]}}^{\gamma\delta} \left ( - \partial_\delta F_{\!\text{[I]}\alpha\beta} +\partial_\alpha F_{\!\text{[I]}\beta\delta} +\partial_\beta F_{\!\text{[I]}\delta\alpha} \right)
\end{align*}
are the Christoffel symbols of the second kind in the curvilinear plane.\\

\begin{figure*}[!h]
\centering
\includegraphics[trim = 0cm 4.5cm 7cm 1cm , clip = true, width=1\textwidth]{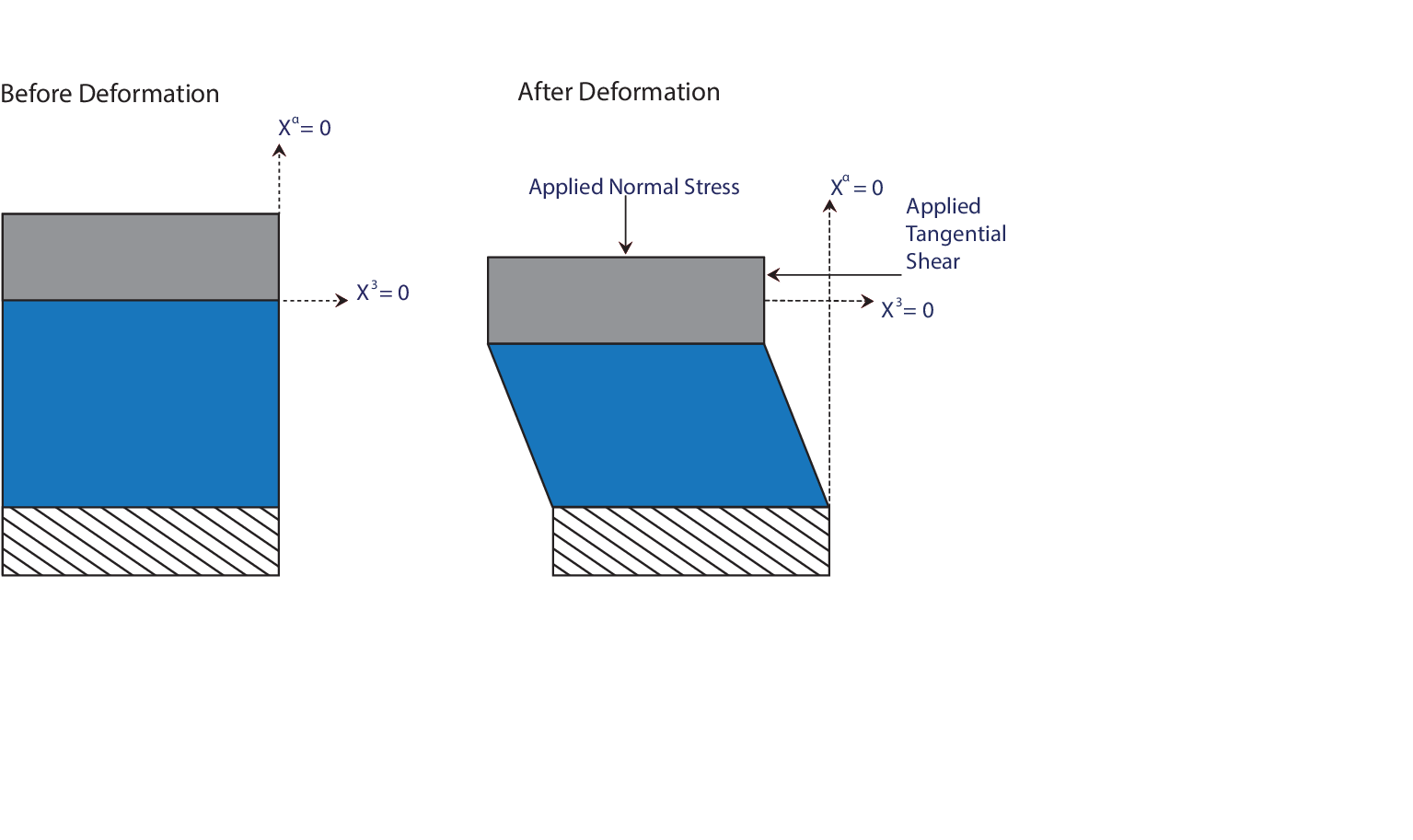}
\caption{Schematic representation of a two-body problem in curvilinear space: a shell (upper body) on an elastic foundation (middle body), before and after deformation}
\label{two-body}
\end{figure*}

Now assume that the shell is coupled to the elastic foundation with friction, where a portion of the foundation is satisfying the zero-Dirichlet boundary condition (i.e. clamped). Also, assume that one is applying forces to both the top and to a portion of the boundary of the shell to mimic compression and shear at the contact region respectively  (see Fig. \ref{two-body}). Now, if the higher the compression, then the higher the normal displacement towards the bottom, i.e. $u^3|_{\omega^+}<0$ (condition \ref{dfnShell} can guarantee this for sensible boundary tractions), and if the higher the shear, then the higher the tangential displacement in the direction of the applied tangential shear, i.e. $\left(u_\alpha u^\alpha\right)^{\frac{1}{2}}|_{\omega^+}>0$, where $\omega^+ = \lim_{x^3\to 0^+} \{\omega \times [0,h)\}$  and by convention $\left(u_\alpha u^\alpha\right)^{\frac{1}{2}} = \sqrt{u_1u^1 + u_2u^2}$. Now, we consider Kikuchi and Oden's model for Coulomb's law of static friction \cite{Kikuchi} for a thin three-dimensional elastic body (i.e. prior to approximating the thin body with a shell), and once extended to curvilinear coordinates and after taking the limit $\varepsilon \to 0$, we find the following
\begin{align}
\left[T^\beta_3(\boldsymbol v)+ \nu_F g_3 (v_\alpha v^\alpha)^{-\frac{1}{2}} v^\beta T_3^3(\boldsymbol v) \right] \!|_{\omega^+} \leq 0 , \label{KikuchiEqn1}
\end{align}
for $T_3^3(\boldsymbol v) |_{\partial\omega^+} < 0$, where $\boldsymbol v$ is the displacement field and the volume $\{\omega \times [0,h]\}$ describes the reference configuration of this elastic body. Just as it is
for Coulomb's friction case (where the bodies are in relative equilibrium given that the magnitude of the normal stress is above a certain factor of the magnitude of the tangential stress), we assume that the bodies (i.e. the thin body and the foundation) are in relative equilibrium given that the normal displacement is a below a certain factor of the magnitude of the tangential displacement, i.e. 
\begin{align}
\left[\left(u_\alpha u^\alpha\right)^{\frac{1}{2}} + C g_3 u^3\right] \!|_{\omega^+} \leq 0,  \label{DisplacementEG}
\end{align}
if $u^3|_{\omega^+} \leq 0$, for some dimensionless constant $C$. To determine the constant $C$, we consider Coulomb's law of static friction for the limiting equilibrium case (i.e. at the point of slipping) and rearrange equation (\ref{KikuchiEqn1}) to obtain the following
\begin{align*}
\bigg[\bar\nabla_{\!3} \left( (v_\alpha v^\alpha)^{\frac{1}{2}} + 2\nu_F \left(1 +\frac{\gamma}{1-2\gamma}\right) g_3  v^3\right) + \left(\frac{2\nu_F\gamma}{1-2\gamma}\right) g_3 \bar\nabla_{\!\alpha} v^\alpha &\\
+ \frac{v^\delta\bar\nabla_{\!\delta} v_3}{(v_\alpha v^\alpha)^{\frac{1}{2}}} & \bigg]|_{\omega^+} = 0.
\end{align*}
Now, the above equation must hold for all elastic conditions, even under extreme conditions such as the \emph{incompressible elasticity} condition,  i.e. $(\frac{\gamma}{1-2\gamma}) \bar\nabla_{\! i} v^i $ $= p(x^1,x^2,x^3)$, where $p(\cdot)$ is a finite function. Thus, we may assume the following equation
\begin{align*}
\bigg[\bar\nabla_{\!3}\left( (v_\alpha v^\alpha)^{\frac{1}{2}} + 2\nu_F g_3 v^3\right)  + \frac{v^\delta\bar\nabla_{\!\delta} v_3}{(v_\alpha v^\alpha)^{\frac{1}{2}}}+ 2\nu_F g_3 p(x^1,x^2,x^3)&\bigg]|_{\omega^+} = 0.
\end{align*}
Now, nondimensionalise the above equation by making the transformations $v^i = \ell w^i$, $x^\alpha = \ell y^\alpha$ and $x^3 = h y^3$ where $\ell = \sqrt{\mathrm{meas}(\boldsymbol \sigma (\omega);\textbf{E}^2)}$, and where $\mathrm{meas}(\cdot;\mathbb{R}^k)$ is standard Lebesgue measure in $\mathbb{R}^k$ (see chapter 6 of Schilling \cite{schilling2017measures}), to obtain the following
\begin{align}
\bigg[ \left(\frac{\ell}{h}\right) & g^3 \frac{\partial}{\partial y^3} \left( (w_\alpha w^\alpha)^{\frac{1}{2}} + 2\nu_F g_3 w^3\right) \nonumber \\
& + \frac{g_3 w^\delta }{(w_\alpha w^\alpha)^{\frac{1}{2}}} \left( \frac{\partial w^3}{\partial y^\delta } + \Gamma^3_{\!\delta i}w^i  \right) 
+ 2\nu_F p(\ell y^1,\ell y^2,hy^3) \bigg]  |_{\omega^+} = 0 \label{CulombsEG2}
\end{align}
where  $g^3 = \sqrt{g^{33}}|_{\omega} = 1$ by construction. As our goal is to study shells, we consider the limit $(h/\ell) \to 0$. As we also require Coulomb's law of static friction for the limiting-equilibrium case  to stay finite in this limit, equation (\ref{CulombsEG2}) implies that
\begin{align*}
\left[ (w_\alpha w^\alpha)^{\frac{1}{2}} + 2\nu_F g_3 w^3\right] \!|_{\{\left(\frac{\omega}{\ell^2}\right) \times [0,1]\}} =  q(y^1,y^2)  + \mathcal{O}\left(\left(\frac{h}{\ell}\right), y^3 \right),
\end{align*}
where $q(\cdot)$ is a finite function. As the above equation must hold true when the bodies are deformation free, we find that $q(\cdot) = 0$. Furthermore, as we are seeking for a relation of the form of equation (\ref{DisplacementEG}), we may assume that $C=2\nu_F$ is a good approximation. Finally, assuming that $\boldsymbol u$ is continuous on $\bar\Omega$, we arrive at the following hypothesis:

\begin{hypothesis} \label{hyp3}
A shell supported by an elastic foundation with a rough contact area that is in agreement with condition \ref{dfnShell} satisfies the following displacement-based friction condition
\begin{align*}
\left[2\nu_F g_3 u^3+ \left(u_\alpha u^\alpha\right)^{\frac{1}{2}}\right]\!|_\omega \leq 0 ,
\end{align*}
where $\nu_F$ is the coefficient of friction between the shell and the foundation, and $\boldsymbol u$ is the displacement field of the shell with respect to the contact region $\omega$. If $[2\nu_F g_3 u^3 + \left(u_\alpha u^\alpha\right)^{\frac{1}{2}}]|_\omega < 0$, then we say that the shell is bonded to the foundation, and if $[2\nu_F g_3 u^3 + \left(u_\alpha u^\alpha\right)^{\frac{1}{2}}]|_\omega =0 $, then we say that the shell is at limiting-equilibrium.
\end{hypothesis}

Using hypothesis \ref{hyp3}, we can now express the energy functional of a shell on a elastic foundation subject to the displacement-based friction condition, and thus, we obtain the following:

\begin{Theorem}\label{thrmShellFriction}
Let $\Omega \subset \mathbb{R}^3$ be a connected open bounded domain that satisfies the segment condition with a uniform-$C^1(\mathbb R^3;\mathbb R^2)$ boundary $\partial\Omega$ such that $\omega, \partial\Omega_0\subset\partial\Omega$ with $\bar\omega\cap\bar{\partial\Omega_0} ={\O}$ and $\mathrm{meas}(\partial\Omega_0;\mathbb{R}^2)>0$, and let $\omega \subset \mathbb{R}^2$ be a connected open bounded plane that satisfies the segment condition with a uniform-$C^1(\mathbb{R}^2;\mathbb{R})$ boundary $\partial\omega$. Also, let $\bar{\boldsymbol X} \in C^2(\bar\Omega;\textbf{E}^3)$ be a diffeomorphism and $\boldsymbol \sigma \in C^3(\bar\omega;\textbf{E}^3)$ be an injective immersion satisfying $0\leq h^2K< hH\ll  1 $ in $\omega$. Furthermore, let $\boldsymbol f \in \boldsymbol L^2(\Omega)$, $\boldsymbol f_{\!0} \in \boldsymbol L^2(\omega)$ (where $f_{\!0}^3\leq 0 ~\mathrm{a.e.}$) and $\boldsymbol \tau_{\!0} \in \boldsymbol L^2(\partial\omega)$ (where $\tau_{\!0}^3\leq 0 ~\mathrm{a.e.}$). Then there exists a unique field $\boldsymbol u \in \boldsymbol V_{\!\!\!\mathscr{F}}(\omega,\Omega)$ such that $\boldsymbol u$ is the solution to the following minimisation problem
\begin{align*}
J(\boldsymbol u) = \min_{\boldsymbol v \in \boldsymbol V_{\!\!\!\mathscr{F}}(\omega,\Omega)} \!\!\! \!\!\! J(\boldsymbol v) ,
\end{align*}
where
\begin{align*}
\boldsymbol V_{\!\!\!\mathscr{F}}(\omega,\Omega) = \{ &\boldsymbol v \in \boldsymbol V_{\!\!\!\mathscr{S}}(\omega,\Omega)\mid \left[2\nu_F g_3 v^3+(v_\alpha v^\alpha)^{\frac{1}{2}}\right]\!|_\omega \leq 0 ~ \mathrm{a.e.} \},\\
\boldsymbol V_{\!\!\!\mathscr{S}}(\omega,\Omega) = \{ &\boldsymbol v \in \boldsymbol H^1(\Omega) \mid \boldsymbol v|_\omega \in H^1(\omega)\!\times\! H^1(\omega)\!\times\! H^2(\omega), \\
& ~~~\boldsymbol v|_{\partial\Omega_0} = \boldsymbol 0,~ \partial_\beta (v^3 |_{\omega})|_{\partial\omega} =0 ~ \beta \in\{1,2\}\},
\end{align*}
\begin{align*}
J(\boldsymbol u) = &\int_{\Omega} \left[\frac{1}{2}A^{ijkl}E_{ij}(\boldsymbol u) E_{kl}(\boldsymbol u) - f^i u_i \right]d\Omega \\
&+\int_{\omega}\bigg[\frac{1}{2}B^{\alpha\beta\gamma\delta}\bigg(h\epsilon_{\alpha\beta}(\boldsymbol u) \epsilon_{\gamma\delta}(\boldsymbol u) + \frac{1}{3}h^3\rho_{\alpha\beta}(\boldsymbol u) \rho_{\gamma\delta} (\boldsymbol u) \bigg) - h f_0^i u_i \bigg]d\omega\\
& - \int_{\partial\omega} h \tau_{\!0}^i u_i ~d(\partial\omega) ,
\end{align*}
and $\nu_F$ is the coefficient of friction between the foundation and the shell.
\end{Theorem}
Note that $L^k(\cdot)$ are the standard $L^k$-Lebesgue spaces and $H^k(\cdot)$ are the standard $W^{k,2}(\cdot)$-Sobolev spaces (see section 5.2.1 of Evans \cite{Evans}), and $\mathrm{a.e.}$ means almost everywhere (see definition 1.40 of Adams and Fournier \cite{adams2003sobolev}). Also note that $f_{\!0}^3\leq 0$ and $\tau_{\!0}^3\leq 0$ respectively represent non-positive normal-force density and non-positive normal-tractions to stay consistent with hypothesis \ref{hyp3}.

\begin{proof}
Note that there exists a unique field $\boldsymbol u \in \boldsymbol V_{\!\!\!\mathscr{S}}(\omega,\Omega)$ such that $\boldsymbol u$ is the solution to the following minimisation problem
\begin{align*}
J(\boldsymbol u) = \min_{\boldsymbol v \in \boldsymbol V_{\!\!\!\mathscr{S}}(\omega,\Omega)} \!\!\! \!\!\! J(\boldsymbol v) ,
\end{align*}
and we refer the reader to section 3.4 of Jayawardana \cite{jayawardana2016mathematical} for the proof. Now, as $(\boldsymbol V_{\!\!\!\mathscr{F}}(\omega,\Omega), J(\cdot)) \subset (\boldsymbol V_{\!\!\!\mathscr{S}}(\omega,\Omega), J(\cdot))$ by construction, it is sufficient to show that $2\nu_F g_3 u^3 + \left(u_\alpha u^\alpha\right)^{\frac{1}{2}} \leq 0$ $\mathrm{a.e.}$ in $\omega$ is a convex functional. \\

Now, let $I(\boldsymbol u; U) = \int_U [2\nu_F g_3 u^3 +\left(u_\alpha u^\alpha\right)^{\frac{1}{2}}]~dx^1dx^2$. By construction $I(\boldsymbol u; U)\leq 0$, $\forall ~U \in \mathcal{M}(\omega)$ with $\mathrm{meas}(U;\omega)>0$, where $\mathcal{M}(\cdot)$ is a $\sigma$-algebra (see definition 1.37 of Adams and Fournier \cite{adams2003sobolev}). Also, by construction $\boldsymbol F_{\!\text{[I]}}$ is positive definite in $\bar \omega$ (see section 5.3 of Kay \cite{kay1988schaum}) and this implies that $I(\cdot; U)$ is a convex functional for all $U \in \mathcal{M}(\omega)$ with $\mathrm{meas}(U;\omega)>0$, i.e.  $I(t \boldsymbol u + (1-t) \boldsymbol v ; U) \leq tI(\boldsymbol u; U) + (1-t)I( \boldsymbol v ; U)$. Furthermore, $I(t \boldsymbol u + (1-t) \boldsymbol v ; U) \leq 0$, $\forall ~U \in \mathcal{M}(\omega)$ with $\mathrm{meas}(U;\omega)>0$, and thus, our convexity result does not violate the definition of the functional $I(\cdot ; U)$, i.e. the condition $2\nu_F g_3 u^3 + \left(u_\alpha u^\alpha\right)^{\frac{1}{2}} \leq 0$ $\mathrm{a.e.}$ in $\omega$ is not violated. Now the proof follows from section 2.6 of  Kinderlehrer and Stampacchia \cite{kinderlehrer2000introduction} or section of 8.4.2 of Evans \cite{Evans}.
\end{proof}

Theorem \ref{thrmShellFriction} implies that there exists a unique weak solution to our  problem. However, due to the free-boundary constraint $[2\nu_F g_3 u^3+\left(u_\alpha u^\alpha\right)^{\frac{1}{2}}]|_\omega \leq 0$ $\mathrm{a.e.}$, the unique minimiser $\boldsymbol u$ may fail to be a critical point in $(\boldsymbol V_{\!\!\!\mathscr{F}}(\omega,\Omega),J(\cdot))$, and thus, one requires the following corollary to find governing equations:

\begin{corollary}\label{crlShell}
Let $\boldsymbol u \in \boldsymbol V_{\!\!\!\mathscr{F}}(\omega,\Omega)$ be  the unique solution to the minimisation problem $J(\boldsymbol u) = \min_{\boldsymbol v \in \boldsymbol V_{\!\!\!\mathscr{F}}(\omega,\Omega)} J(\boldsymbol v)$, then we get the following variational inequality
\begin{align*}
0\leq J^\prime (\boldsymbol u)(\boldsymbol v -\boldsymbol u),~\forall~\boldsymbol v \in \boldsymbol V_{\!\!\!\mathscr{F}}(\omega,\Omega).
\end{align*}
\end{corollary}
\begin{proof}
$(\boldsymbol V_{\!\!\!\mathscr{F}}(\omega,\Omega), J(\cdot))$ is a convex space, and thus, the proof follows from section 8.4.2 of Evans \cite{Evans}. 
\end{proof}

\subsection{The Equations of Equilibrium} 

We assume that $\boldsymbol u \in C^2(\Omega;\mathbb{R}^3)$, $u^\beta|_{\omega} \in C^3(\omega)$, $u^3|_{\omega} \in C^4(\omega)$ and $ 2\nu_F g_3 u^3+\left(u_\alpha u^\alpha\right)^{\frac{1}{2}}\leq 0$ everywhere in $\omega$, and thus, theorem \ref{thrmShellFriction}, corollary \ref{crlShell} and the \emph{principle of virtual displacements}  (see section 2.2.2 of Reddy \cite{reddy2006theory}) imply that the \emph{governing equations of the elastic foundation} can be expressed as follows
\begin{align*}
\bar\nabla_{\!i} T^i_j(\boldsymbol u) + f_j & =0,  ~ j\in\{1,2,3\},
\end{align*}
and the \emph{boundary conditions of the elastic foundation} can be expressed as follows
\begin{align*}
\boldsymbol u|_{\partial\Omega_0} &= \boldsymbol 0,\\
[\bar n_i T^i_j(\boldsymbol u)]|_{\partial\Omega\setminus \{\omega \cup \partial\Omega_0\}}&= 0,  ~ j\in\{1,2,3\},
\end{align*}
where $\bar{\boldsymbol n}$ is the unit outward normal to the boundary $\partial\Omega$ in curvilinear coordinates.\\

As for the \emph{governing equations of the frictionally coupled shell}, notice that the set $\boldsymbol V_{\!\!\!\mathscr{F}}(\omega,\Omega)$ is not a linear set as it violates the homogeneity property. However, it can be shown that for any field $  \boldsymbol u \in \boldsymbol V_{\!\!\!\mathscr{F}}(\omega,\Omega)$, there exists a field $  \boldsymbol w \in \boldsymbol V_{\!\!\!\mathscr{F}}(\omega,\Omega)\!\setminus \!\{\boldsymbol u\}$ and a constant $\varepsilon >0$ such that $ \boldsymbol u+ s\boldsymbol w \in \boldsymbol V_{\!\!\!\mathscr{F}}(\omega,\Omega)$, $\forall ~s \in (-\varepsilon,1]$, i.e. $\int_U [2\nu_F g_3 (u^3+sw^3) +(u_\alpha u^\alpha+2su_\alpha w^\alpha + s^2 w_\alpha w^\alpha)^{\frac{1}{2}}]~dx^1dx^2\leq0$, $\forall~ U \in \mathcal{M}(\omega)$ with $\mathrm{meas}(U;\omega)>0$, which we show as follows.\\

To find the governing equations for the $[2\nu_F g_3 u^3 + \left(u_\alpha u^\alpha\right)^{\frac{1}{2}}]|_{\omega} < 0$ case, consider a unique minimiser $\boldsymbol u \in \boldsymbol  V_ {\!\!\!\mathscr{O}}(\omega,\Omega)$, where $\boldsymbol V_ {\!\!\!\mathscr{O}}(\omega,\Omega) = \{\boldsymbol v \in \boldsymbol V_{\!\!\!\mathscr{F}}(\omega,\Omega)\mid [2\nu_F g_3 v^3 + (v_\alpha v^\alpha)^{\frac{1}{2}}]|_{\omega_ {\mathscr{O}}} < 0 ~ \mathrm{a.e.}\}$ and where $\omega_ {\mathscr{O}} = \{V \in \mathcal{M}(\omega) \mid [2\nu_F g_3 v^3 + (v_\alpha v^\alpha)^{\frac{1}{2}}]|_V < 0 ~ \mathrm{a.e.},~ \mathrm{meas}(V;\omega)>0\}$. Now, given a $\boldsymbol w \in \boldsymbol  V_ {\!\!\!\mathscr{O}}(\omega,\Omega)$, there exists an $\varepsilon>0$ such that we get $\boldsymbol u + s\boldsymbol w \in \boldsymbol V_{\!\!\!\mathscr{F}}(\omega,\Omega)$, $\forall ~s \in (-\varepsilon,1]$ where $$\varepsilon < \frac{\left(2\nu_F ||u^3||_{L^1(U)}- ||(u_\gamma u^\gamma)^{\frac{1}{2}}||_{L^1(U)}\right)}{\left(2\nu_F ||w^3||_{L^1(U)}+ ||(w_\alpha w^\alpha)^{\frac{1}{2}}||_{L^1(U)}\right)}$$ for some $U \in \mathcal{M}(\omega_{\mathscr{O}})$. Now, simply let $\boldsymbol v = \boldsymbol u + s\boldsymbol w$ in corollary 4 to obtain the inequality $0\leq J^\prime (\boldsymbol u)(s\boldsymbol w)$, $\forall ~s \in (-\varepsilon,1]$ for this $\boldsymbol w \in  \boldsymbol  V_ {\!\!\!\mathscr{O}}(\omega,\Omega)$. Finally, noticing that $0\leq \mathrm{sign}(s) J^\prime (\boldsymbol u)( |s|\boldsymbol w)   $, $\forall~ \boldsymbol w \in \boldsymbol  V_ {\!\!\!\mathscr{O}}(\omega,\Omega)$, we get the governing equations for the \emph{bonded} case:\\
If $\big[2\nu_F g_3 u^3 + \left(u_\alpha u^\alpha\right)^{\frac{1}{2}}\big]|_{\omega}< 0$, then $\boldsymbol G(\boldsymbol u) = \boldsymbol 0$, where
\begin{align*}
G_\beta (\boldsymbol u) = ~& \nabla_{\!\alpha} \tau^\alpha_\beta(\boldsymbol u) + \frac{2}{3}h^2F_{\!\text{[II]}\beta}^{~~\alpha}\nabla_{\!\gamma} \eta^\gamma_\alpha(\boldsymbol u) \\
&+ \frac{1}{3}h^2\left(\nabla_{\!\gamma}F_{\!\text{[II]}\beta}^{~~\alpha}\right) \eta^\gamma_\alpha(\boldsymbol u)  - \frac{1}{h}\mathrm{Tr}(T^3_\beta(\boldsymbol u)) + f_{0\beta}   ,~ \beta\in\{1,2\},\\
G_3 (\boldsymbol u) =~& g_3 \Big[F_{\!\text{[II]}\alpha}^{~~\gamma}\tau^\alpha_\gamma(\boldsymbol u)  - \frac{1}{3}h^2\nabla_{\!\alpha} \left(\nabla_{\!\gamma}\eta^{\alpha\gamma}(\boldsymbol u)\right)\\
& ~~~+ \frac{1}{3}h^2F_{\!\text{[II]}\alpha}^{~~\delta}F_{\!\text{[II]}\gamma}^{~~\alpha} \eta^\gamma_\delta(\boldsymbol u) - \frac{1}{h}\mathrm{Tr}(T^3_3(\boldsymbol u))\Big] + f_{03} , ~(\text{where} f_{\!03}\leq 0),
\end{align*}
and where $\mathrm{Tr}(T^3_j(\boldsymbol u)) = T^3_j(\boldsymbol u)|_\omega$ and $\mathrm{Tr}(\cdot)$ is the trace operator (see section 5.5 of Evans \cite{Evans}).\\

To find the governing equations for the $[2\nu_F g_3 u^3 + \left(u_\alpha u^\alpha\right)^{\frac{1}{2}}]|_{\omega} = 0$ case, consider a unique minimiser $\boldsymbol u \in \boldsymbol  V_ {\!\!\!\mathscr{C}}(\omega,\Omega)$, where $\boldsymbol V_ {\!\!\!\mathscr{C}}(\omega,\Omega) = \{ \boldsymbol v \in \boldsymbol V_{\!\!\!\mathscr{F}}(\omega,\Omega)\mid  [2\nu_F g_3 v^3 + (v_\alpha v^\alpha)^{\frac{1}{2}}]|_{\omega_ {\mathscr{C}}} = 0 ~ \mathrm{a.e.}\}$ and where $\omega_ {\mathscr{C}} = \{V \in \mathcal{M}(\omega) \mid [2\nu_F g_3 v^3 + (v_\alpha v^\alpha)^{\frac{1}{2}}]|_V = 0 ~ \mathrm{a.e.},~ \mathrm{meas}(V;\omega)>0\}$. Now, noticing that $u^j|_{\omega_ {\mathscr{C}}}$ are not independent, but are related  by  the condition $u^3 |_{\omega_ {\mathscr{C}}} = -\frac12 \nu_F^{-1}g^3\left(u_\alpha u^\alpha\right)^{\frac{1}{2}}|_{\omega_ {\mathscr{C}}}$, we get $ \delta u^3 |_{\omega_ {\mathscr{C}}} =  - \frac{1}{2} \nu_F^{-1} g^3 [\left(u_\alpha u^\alpha\right)^{-\frac{1}{2}} (u_\gamma \delta u^\gamma)]|_{\omega_ {\mathscr{C}}}$. Now let
\begin{align*}
\boldsymbol V_ {\!\!\!\mathscr{C}}(\boldsymbol u; \omega,\Omega) = \{~&\boldsymbol v \in \boldsymbol V_ {\!\!\!\mathscr{C}}(\omega,\Omega) \mid \\
& \boldsymbol (v^1, v^2\boldsymbol)|_{\omega_ {\mathscr{C}}} = \boldsymbol (c u^1, c u^2\boldsymbol)|_{\omega_ {\mathscr{C}}}~ \mathrm{a.e.},~  \forall c >0, ~ \boldsymbol u \in \boldsymbol V_ {\!\!\!\mathscr{C}}(\omega,\Omega)\},
\end{align*}
and thus, given a $ \boldsymbol w \in \boldsymbol V_ {\!\!\!\mathscr{C}}(\boldsymbol u; \omega,\Omega)$ there exists an $\varepsilon>0$  such that we get $\boldsymbol u + s\boldsymbol w \in \boldsymbol V_{\!\!\!\mathscr{F}}(\omega,\Omega) $, $\forall ~s \in (-\varepsilon,1]$, where $\varepsilon <  ||\left(w_\alpha w^\alpha\right)^{\frac{1}{2}}||_{L^1(U)}^{-1}||\left(u_\gamma u^\gamma\right)^{\frac{1}{2}}||_{L^1(U)}$ for some $U \in \mathcal{M}(\omega_{\mathscr{C}})$. Now, simply let $\boldsymbol v = \boldsymbol u + s\boldsymbol w$ in corollary 4 to obtain $0\leq J^\prime (\boldsymbol u)(s\boldsymbol w|_\Omega + s\boldsymbol( w^1, w^2 \boldsymbol)|_{\omega_ {\mathscr{C}}})$, $\forall ~s \in (-\varepsilon,1]$ for this $ \boldsymbol w \in \boldsymbol V_ {\!\!\!\mathscr{C}}(\omega,\Omega)$. Finally, noticing that $ J^\prime (\boldsymbol u)(\boldsymbol w|_\Omega)$ $= 0 $ (this leads to the governing equations in the foundation) and  $0\leq\mathrm{sign}(s)  J^\prime (\boldsymbol u)(  |s|\boldsymbol( w^1, w^2 \boldsymbol)|_{\omega_ {\mathscr{C}}} ),~ \forall~ \boldsymbol w  \in \boldsymbol V_ {\!\!\!\mathscr{C}}(\boldsymbol u; \omega,\Omega) \subset \boldsymbol V_ {\!\!\!\mathscr{C}}( \omega,\Omega)$, we get the governing equations for the \emph{limiting-equilibrium} case (adapted from section 8.4.2 of Evans \cite{Evans}):\\
If $\big[2\nu_F g_3 u^3 + \left(u_\alpha u^\alpha\right)^{\frac{1}{2}}\big]|_{\omega} = 0$, then
\begin{align*}
 G^\beta(\tilde{\boldsymbol u})  - \frac{1}{2}\frac{g_3}{\nu_F} \frac{\tilde u^\beta}{\left(\tilde u_\alpha \tilde u^\alpha \right)^{\frac{1}{2}}} G^ 3 (\tilde{\boldsymbol u}) = 0, ~ \beta\in\{1,2\},
\end{align*}
where 
\begin{align*}
\tilde{\boldsymbol u} & = \boldsymbol ( u^1,u^2, - \frac{1}{2}\frac{g^3}{\nu_F}\left(u_\alpha u^\alpha\right)^{\frac{1}{2}} \boldsymbol)|_\omega ,\\
\boldsymbol ( \partial_3 \tilde u^1,\partial_3 \tilde u^2,\partial_3 \tilde u^3 \boldsymbol) & = \boldsymbol ( \partial_3 u^1,\partial_3 u^2,\partial_3 u^3 \boldsymbol)|_\omega .
\end{align*}

Finally, the \emph{boundary conditions of the frictionally coupled shell} can be expressed as follows
\begin{align*}
\big[n_\alpha\tau^\alpha_\beta(\boldsymbol u) + \frac{2}{3}h^2 n_\gamma F_{\!\text{[II]}\beta}^{~~\alpha} \eta^\gamma_\alpha(\boldsymbol u)\big] |_{\partial \omega} & = \tau_{\!0\beta} , ~ \beta\in\{1,2\}, \\
- \frac{1}{3}h^2 g_3 n _\gamma \nabla_{\!\alpha}\eta^{\alpha\gamma}(\boldsymbol u)  |_{\partial \omega} & = \tau_{\!0 3}, ~(\text{where} ~\tau_{\!03}\leq 0),\\
\partial_\beta u^3 |_{\partial\omega} & = 0,~ \beta\in\{1,2\},
\end{align*}
where $\boldsymbol n$ is the unit outward normal vector to the boundary $\partial\omega$ in curvilinear coordinates and $\boldsymbol \tau_{\!0}$ is an external traction field acting on the boundary of the frictionally coupled shell.

\subsection{A Numerical Example}
\label{ch4Num}

\begin{figure}[!h]
\centering
\includegraphics[trim = 0cm 0.5cm 0cm 0cm , clip = true, width=0.75\linewidth]{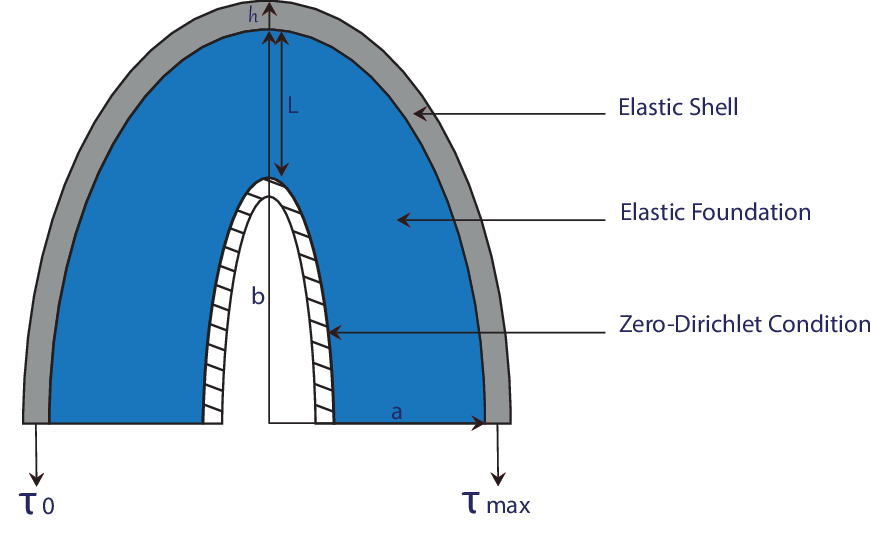}
\caption{Schematic representation of a shell frictionally coupled to an elastic foundation, where the cross-section of the contact region forms a semi-ellipse}
\label{shellellipse}
\end{figure}

Assume that we are dealing with an overlying shell with a thickness $h$ that is frictionally coupled to an elastic foundation, where the unstrained configuration of the foundation is an infinitely long annular semi-prism parametrised by the following diffeomorphism 
\begin{align*}
\bar {\boldsymbol X}(x^1,x^2,x^3) =~&\boldsymbol(x^1, ~a\sin(x^2),~b\cos(x^2)\boldsymbol)_\text{E} + \frac{x^3}{\varphi(x^2)}\boldsymbol (0,~b\sin(x^2), ~a\cos(x^2)\boldsymbol )_\text{E},
\end{align*}
where $\varphi (x^2)= (b^2\sin^2(x^2) + a^2\cos^2(x^2))^{\frac{1}{2}}$, $ x^1 \in (-\infty,\infty)$, $x^2 \in (-\frac{1}{2}\pi,\frac{1}{2}\pi)$, $x^3 \in (-L,0)$, and $a$ is the horizontal radius and $b$ is the vertical radius of the contact region (see Fig. \ref{shellellipse}). Thus, the equations of the foundation can be expressed as follows
\begin{align*}
(\bar\lambda + \bar\mu)\partial_2(\bar \nabla_{\!i} u^i ) + \bar \mu (\bar\psi_2)^2 \bar \Delta u^2 & = 0,\\
(\bar \lambda + \bar \mu)\partial_3(\bar \nabla_{\!i} u^i) + \bar \mu \bar \Delta u^3 & = 0,
\end{align*}
where $\boldsymbol u = \boldsymbol(0,u^2(x^2,x^3),u^3(x^2,x^3) \boldsymbol )$ is the displacement field, $\bar\Delta = \bar\nabla_i\bar\nabla^i $ is the vector-Laplacian operator in the curvilinear space (see page 3 Moon and Spencer \cite{moon2012field}) with respect to $\Omega^\text{New} =  (-\frac{1}{2}\pi,\frac{1}{2}\pi) \times(-L,0)$ and $ \bar\psi_2 = \varphi (x^2)+ x^3 ab(\varphi(x^2))^{-2}$.\\

Now, eliminating $x^1$ dependency, one can express the remaining boundaries as follows
\begin{align*}
\partial \Omega^\text{New} & = \bar \omega^\text{New}\cup \partial \Omega_0^\text{New} \cup \partial \Omega_f^\text{New},\\
\omega^\text{New} & = (-\frac{1}{2}\pi,\frac{1}{2}\pi)\times \{0\} ,\\
\partial \Omega_0^{\text{New}} & = (-\frac{1}{2}\pi,\frac{1}{2}\pi)\times \{-L\} ,\\
\partial \Omega_f^{\text{New}} &= \{\{-\frac{1}{2}\pi\}\times (-L,0)\}\cup \{\{\frac{1}{2}\pi\}\times (-L,0)\}.
\end{align*}
Thus, the boundary conditions that one imposes on the foundation reduce to the following
\begin{align*}
u^2|_{\overline{\partial \Omega}_0^\text{New}} &= 0~\text{(zero-Dirichlet)},\\
u^3|_{\overline{\partial \Omega}_0^\text{New}} &= 0~\text{(zero-Dirichlet)},\\
\big[(\bar\psi_2)^2\partial_3u^2 +\partial_2u^3 \big]|_{\partial \Omega_f^\text{New}} & = 0~\text{(zero-Robin)},\\
\big[(\bar\lambda + 2\bar\mu) \partial_2 u^2 + \bar\lambda\left( \partial_3 u^3 + \bar\Gamma^2_{\!22} u^2 + \bar\Gamma^2_{\!23} u^3\right)\big] |_{\partial \Omega_f^\text{New}} & = 0~\text{(zero-Robin)}.
\end{align*}

Now, consider the overlying shell's unstrained configuration, which is described by the injective immersion $\boldsymbol \sigma(x^1,x^2) =\boldsymbol (x^1,a\sin(x^2),b\cos(x^2)\boldsymbol )_\text{E}$, where $x^1 \in (-\infty,\infty)$ and $x^2 \in (-\frac{1}{2}\pi,\frac{1}{2}\pi)$. Thus, one can express the governing equations of the shell as follows:\\
If $[2\nu_Fu^3 +\psi_2|u^2|]|_{\omega^\text{New}}<0$, then
\begin{align*}
h \Lambda \partial_2 \epsilon^2_2(\boldsymbol u) + \frac{1}{3}h^3\Lambda (2F_{\!\text{[II]}2}^{~~2}\partial_2 \rho^2_2(\boldsymbol u)  + \partial_2F_{\!\text{[II]}2}^{~~2} \rho^2_2(\boldsymbol u)) - \mathrm{Tr} (T^3_2(\boldsymbol u))  &= 0,\\
- h \Lambda F_{\!\text{[II]}2}^{~~2} \epsilon^2_2(\boldsymbol u) + \frac{1}{3}h^3 \Lambda ( \Delta \rho^2_2(\boldsymbol u)  - F_{\!\text{[II]}2}^{~~2}F_{\!\text{[II]}2}^{~~2} \rho^2_2(\boldsymbol u)) + \mathrm{Tr} (T^3_3(\boldsymbol u))  &= 0,
\end{align*}
where $\boldsymbol u |_{\bar\omega^\text{New}} = \boldsymbol (0,u^2(x^2,0),u^3(x^2,0)\boldsymbol ) $ is the displacement field of the shell, $\Delta = \nabla_\alpha \nabla^\alpha$ is the vector-Laplacian in curvilinear plane (see page 3 Moon and Spencer \cite{moon2012field}) with respect to $\omega^\text{New}= (-\frac{1}{2}\pi,\frac{1}{2}\pi)$ and $ \psi_2 = \varphi(x^2)$;\\
If $[2\nu_Fu^3 +\psi_2|u^2|]|_{\omega^\text{New}}=0$, then
\begin{align*} 
& \nu_F h \Lambda \partial_2 \epsilon^2_2(\tilde{\boldsymbol u})  - \frac{1}{2} h \Lambda \psi_2\mathrm{sign}(\tilde u^2) F_{\!\text{[II]}2}^{~~2} \epsilon^2_2 (\tilde{\boldsymbol u}) \nonumber\\
& ~~~+\frac{1}{3}\nu_F h^3 \Lambda (2F_{\!\text{[II]}2}^{~~2}\partial_2 \rho^2_2(\tilde{\boldsymbol u}) + \partial_2F_{\!\text{[II]}2}^{~~2} \rho^2_2(\tilde{\boldsymbol u})) \nonumber\\
& ~~~+ \frac{1}{6}h^3 \Lambda \psi_2 \mathrm{sign}(\tilde u^2) (\Delta \rho^2_2 (\tilde{\boldsymbol u}) - F_{\!\text{[II]}2}^{~~2}F_{\!\text{[II]}2}^{~~2} \rho^2_2(\tilde{\boldsymbol u})) \nonumber\\
& ~~~- \nu_F \mathrm{Tr} (T^3_2(\tilde{\boldsymbol u})) + \frac{1}{2}\psi_2\mathrm{sign}(\tilde u^2) \mathrm{Tr} (T^3_3(\tilde{\boldsymbol u})) = 0,
\end{align*}
where
\begin{align*}
\tilde {\boldsymbol u} &= \boldsymbol ( 0,u^2, -\frac12 \nu_F^{-1}\psi_2|u^2|\boldsymbol)|_{\omega^\text{New}},\\
\boldsymbol ( 0,\partial_3 \tilde u^2, \partial_3 \tilde u^3\boldsymbol) & = \boldsymbol ( 0,\partial_3 u^2, \partial_3 u^3\boldsymbol)|_{\omega^\text{New}},
\end{align*}
and where
\begin{align*}
\mathrm{Tr} (T^3_2(\boldsymbol u))  &=\bar\mu\left( (\psi_2)^2 \partial_3u^2 +\partial_2u^3 \right) |_{\omega^\text{New}},\\
\mathrm{Tr} (T^3_3(\boldsymbol u))  &=\big[\bar\lambda\left( \partial_2 u^2 + \bar\Gamma^2_{\!22} u^2 + \bar\Gamma^2_{\!23} u^3\right) + (\bar\lambda + 2\bar\mu) \partial_3 u^3\big]|_{\omega^\text{New}},
\end{align*}
and
\begin{align*}
\Lambda = 4\mu\frac{\lambda+\mu}{\lambda+2\mu}.
\end{align*}

Now, eliminating $x^1$ dependency, one can express the remaining boundaries as follows
\begin{align*}
\partial \omega^\text{New} &= \partial \omega_{T_0}^\text{New}\cup \partial \omega_{T_\text{max}}^\text{New},\\
\partial \omega_{T_0}^\text{New} &= \{0\},\\
\partial \omega_{T_\text{max}}^\text{New} &= \{\pi\}.
\end{align*}
Thus, the boundary conditions of the shell reduce to the following form
\begin{align*}
\big[\Lambda \epsilon^2_2(\boldsymbol u) + \frac23h^2 \Lambda F_{\!\text{[II]}2}^{~~2} \rho^2_2(\boldsymbol u) \big]|_{\{\{\partial \omega_{T_0}^\text{New}, \partial \omega_{T_\text{max}}^\text{New}\}\times [0,h]\}} 
&= \{\tau_0 , \tau_\text{max}\}  ~\text{(traction)} ,\\
\partial_2 \rho^2_2 (\boldsymbol u)|_{\partial \omega^\text{New} } & = 0 ~\text{(zero-pressure)} ,\\
\partial_2 u^3 |_{\partial \omega^\text{New} } &= 0 ~\text{(zero-Neumann)}.
\end{align*}

Despite the fact that the original problem is three-dimensional, it is now a two-dimensional problem as the domain now resides in the set $\{(x^2,x^3) \mid (x^2,x^3)\in [-\frac{1}{2}\pi,\frac{1}{4}\pi]\times[-L,0]\}$. To conduct numerical experiments, we use the second-order accurate fourth-order derivative iterative-Jacobi finite-difference method. Although we use a rectangular grid for discretisation, as a result of the curvilinear nature of the governing equations, there exists an implicit grid dependence implying that the condition $ \psi_0 \Delta x^2 \leq \Delta x^3$, $\forall ~\psi_0 \in \{\bar\psi_2(x^2,x^3) \mid x^2\in[-\frac{1}{2}\pi,\frac{1}{2}\pi] ~\text{and}~ x^3\in[-L,0] \}$ must be satisfied, where $\Delta x^j$ is a small increment in $x^j$ direction in this context. For our purposes, we let $\Delta x^2 = \frac{1}{N-1}\pi$ and $\psi_0 = \bar\psi_2(\frac{1}{4}\pi,0)$, where $N=250$. We also keep the values $ a = 2$ (units: m), $L=1$ (units: m), $\bar E = 10^3$ (units: Pa), $\bar \nu = \frac{1}{4}$ and $\tau_0=1$ (units: N/$\text{m}^2$) fixed for all experiments. \\

\begin{figure*}[!h]
\centering
\includegraphics[trim = 2cm 1cm 2cm 1cm, clip = true, width=1\linewidth]{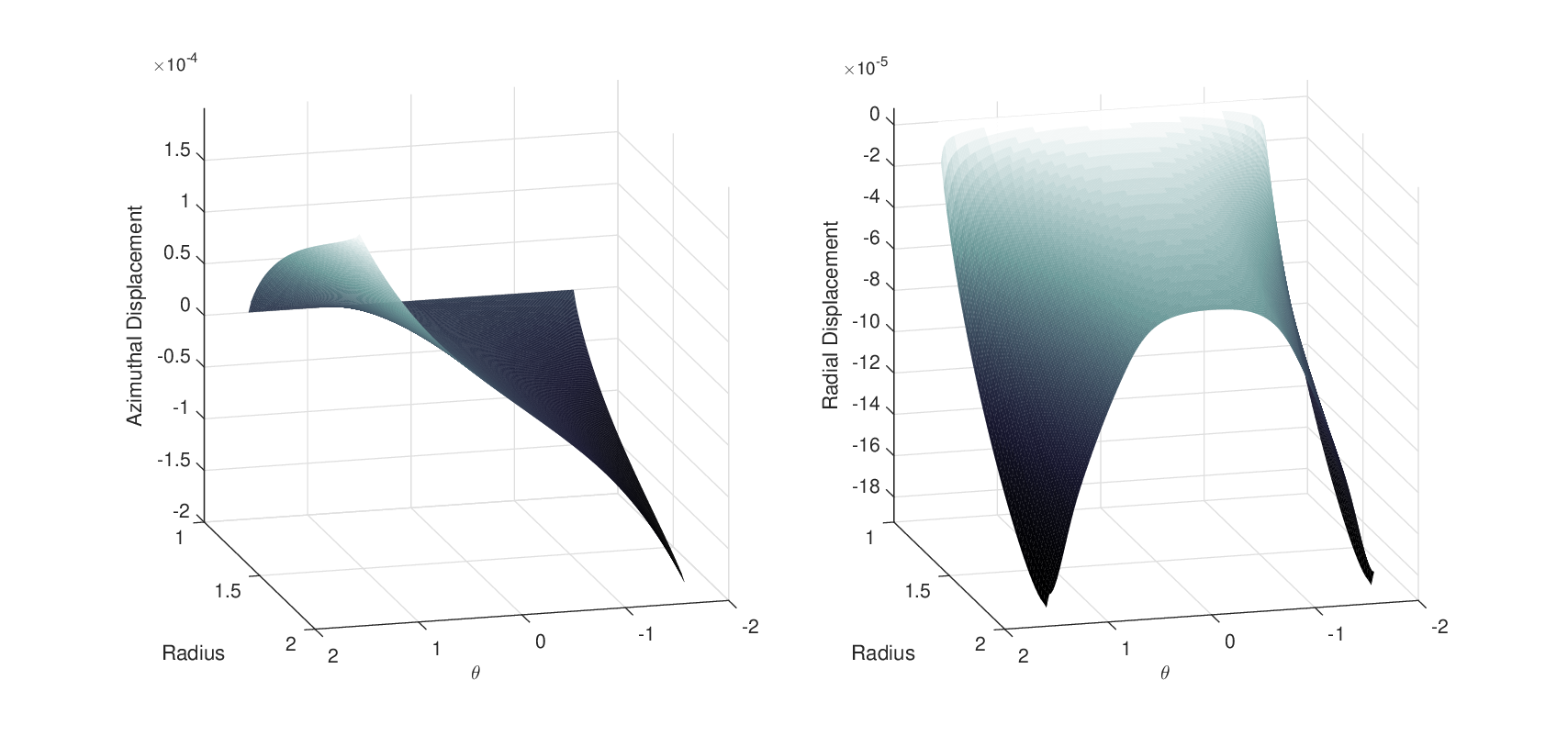}
\caption{Displacement field of the foundation predicted by the shell on an elastic foundation model with friction, where $\theta= x^2$\label{Ch4Shell}}
\end{figure*}

Fig. \ref{Ch4Shell} is calculated with the values of $\tau_\text{max}= 1$N/$\text{m}^2$, $b=2$m, $h=\frac{1}{8}$m, $ E = 8000$Pa, $\nu = \frac{1}{4}$ and $\nu_F = 1$, and it shows the azimuthal (i.e $u^2$) and the radial (i.e. $u^3$) displacements. The maximum azimuthal displacements are observed at $x^2=\pm\frac{1}{2}\pi$, with respective azimuthal displacements of $u^2 = \pm 1.79\times 10^{-4}$rad. The maximum radial displacement is observed at $x^2=\pm \frac{1}{2}\pi$, with a radial displacement of $u^3 =-1.84\times 10^{-4}$m. Furthermore, in the intervals $x^2\in(-1.26,-0.788)$ and $x^2\in(0.788, 1.26)$, we see that the shell is at limiting-equilibrium. These observations simply imply that the shell is more likely to debond from the foundation at the boundaries where we apply external stresses $\tau_0$ and $\tau_\text{max}$, and more likely to stay bonded away from the boundary of the shell. Note that all numerical codes are available at \href{http://discovery.ucl.ac.uk/id/eprint/1532145}{http://discovery.ucl.ac.uk/id/eprint/1532145}.

\section{The Two-Body Coulomb's Law of Static Friction}

The most comprehensive mathematical study on friction that we are aware of is the publication by Kikuchi and Oden \cite{Kikuchi}. Therefore, in this section, we extend their model to study a two-body friction problem in curvilinear coordinates, and we do so by modifying equation (\ref{odenJepsi}) (see section 5.5 (v) of Kikuchi and Oden \cite{Kikuchi}). Assume that an elastic body on a rough rigid surface where the friction is governed by Coulomb's law of static friction. Given that one is using curvilinear coordinates, fix the purely-normal stress at the contact boundary as a constant, i.e. $\nu_F T^3_3(\boldsymbol v)|_{\omega^+} = \mathscr{K}$. Then, equation (\ref{odenJepsi})  implies that $j_\varepsilon^\prime(\boldsymbol v)\delta\boldsymbol v = -\int_\omega g_3 T^3_\alpha(\boldsymbol v)\delta v^\alpha|_{\omega^+} ~d\omega$. Note that $\boldsymbol v|_{\omega^+}$ describes the relative displacement between the elastic body and the boundary $\omega$, and thus, if the elastic body is in contact with another rough elastic body, then the displacement field one must consider is the relative displacement (due to the fact that friction opposes potential relative motion). Now, consider a two-body contact problem where the contact area is rough and the friction is governed by Coulomb's law of static friction. Now, let the displacement fields of the overlying body and the foundation be $\boldsymbol v$ and $\boldsymbol u$ respectively. As the purely-normal stress is continuous at the boundary, just as before, fix the purely normal stress as $\nu_F T^3_3(\boldsymbol u)|_{\omega^-}=\nu_F T^3_3(\boldsymbol v) |_{\omega^+}= \mathscr{K}$, where $\omega^- = \lim_{x^3\to 0^-}\Omega $, and make the transformation  $v^\beta|_{\omega^+} \to v^\beta|_{\omega^+}-u^\beta|_{\omega^-}$ in the functional $j_\varepsilon(\cdot)$ to signify the relative displacement field. Now, collecting all the tangential terms from the contact boundary (i.e. $u^\beta|_{\omega^-}$ and $v^\beta|_{\omega^+}$ terms), one finds $j_\varepsilon^\prime(\boldsymbol v -\boldsymbol u)(\delta\boldsymbol v -\delta\boldsymbol u) = -\int_\omega g_3 ([T^3_\alpha(\boldsymbol v)\delta v^\alpha]|_{\omega^+} -[T^3_\alpha(\boldsymbol u)\delta u^\alpha]|_{\omega^-} ) ~d\omega$, where
\begin{align*}
 \qquad j_\varepsilon (\boldsymbol{v}-\boldsymbol{u}) & = \left\{\begin{aligned}
        \int_\omega\left [\mathscr{K}\Phi(\boldsymbol v -\boldsymbol u)  - \frac{1}{2}\varepsilon\right] d\omega & ,~\text{if}~\Phi(\boldsymbol v -\boldsymbol u) |_\omega \geq \varepsilon,\\
       \int_\omega \left[\frac{1}{2} \mathscr{K} \frac{\left(\Phi(\boldsymbol v -\boldsymbol u)\right) ^2}{\varepsilon} \right] d\omega &,~\text{if}~\Phi(\boldsymbol v -\boldsymbol u) |_\omega< \varepsilon,
       \end{aligned}
 \right.\\
\Phi(\boldsymbol v -\boldsymbol u) & = \left(\ell_\alpha(\boldsymbol v-\boldsymbol u)\ell^\alpha(\boldsymbol v-\boldsymbol u) \right)^{\frac12},\\
\boldsymbol \ell(\boldsymbol v-\boldsymbol u) & = \boldsymbol (v^1, v^2\boldsymbol) |_{\omega^+} -\boldsymbol (u^1, u^2\boldsymbol) |_{\omega^-}.
\end{align*}

As the two bodies are in contact, the normal displacement (of both bodies) is continuous. Thus, one obtains the modified Kikuchi and Oden's model for Coulomb's law of static friction for a two-body problem in curvilinear coordinates, which is described by the following set of equations
\begin{align*}
v^3 |_{\omega^+}-u^3|_{\omega^-} &= 0,\\
T_3^3(\boldsymbol v) |_{\omega^+}- T_3^3(\boldsymbol u) |_{\omega^-}&= 0,
\end{align*}
\begin{align*} 
 \qquad T_3 ^\beta(\boldsymbol v) |_{\omega^+} =  
\left\{\begin{aligned}
        -\frac{\nu_F g_3 \left(v^\beta|_{\omega^+}-u^\beta|_{\omega^-}\right)}{\Phi(\boldsymbol v -\boldsymbol u) }\left(T_3^3(\boldsymbol v) |_{\omega^+}\right) & ,~\text{if}~\Phi(\boldsymbol v -\boldsymbol u)  |_{\omega^+}\geq \varepsilon ,\\
       - \frac{\nu_F g_3 \left(v^\beta|_{\omega^+}-u^\beta|_{\omega^-}\right)}{\varepsilon }\left(T_3^3(\boldsymbol v)  |_{\omega^+}\right)&,~\text{if}~\Phi(\boldsymbol v -\boldsymbol u) |_{\omega^+}< \varepsilon,
       \end{aligned}
 \right.
\end{align*}
\begin{align*} 
 \qquad T_3 ^\beta(\boldsymbol u) |_{\omega^-} = 
\left\{\begin{aligned}
        -\frac{\nu_F g_3 \left(v^\beta|_{\omega^+}-u^\beta|_{\omega^-}\right)}{\Phi(\boldsymbol v -\boldsymbol u) }\left(T_3^3(\boldsymbol u) |_{\omega^-} \right)& ,~\text{if}~\Phi(\boldsymbol v -\boldsymbol u)  |_{\omega^-}\geq \varepsilon ,\\
        -\frac{\nu_F g_3 \left(v^\beta|_{\omega^+}-u^\beta|_{\omega^-}\right)}{\varepsilon }\left(T_3^3(\boldsymbol u)  |_{\omega^-}\right)&,~\text{if}~\Phi(\boldsymbol v -\boldsymbol u) |_{\omega^-}< \varepsilon,
       \end{aligned}
 \right.
\end{align*}
where
\begin{align*}
T^\beta_3(\boldsymbol v) &= \mu\left(\bar\nabla^\beta v_3 + \bar\nabla_{\!3} v^\beta\right ),\\
T^3_3(\boldsymbol v )& = \lambda \bar\nabla_{\!\alpha} v^\alpha + \left(\lambda + 2\mu\right)\bar \nabla_{\!3} v^3,\\
T^\beta_3(\boldsymbol u) &= \bar \mu\left(\bar\nabla^\beta u_3 + \bar\nabla_{\!3} u^\beta\right ),\\
T^3_3(\boldsymbol u )& = \bar \lambda \bar\nabla_{\!\alpha} u^\alpha + \left(\bar\lambda + 2\bar\mu\right)\bar \nabla_{\!3} u^3,
\end{align*}
and where $\nu_F$ is the coefficient of friction. Note that given that $T^3_3(\boldsymbol u )|_{\omega^-}$ is fixed as a positive constant and considering Euclidean coordinates, then the above problem simply reduces to the Kikuchi and Oden's original model for Coulomb's law of static friction \cite{Kikuchi} in the limit $\boldsymbol u \to 0$. Also, the above modified Kikuchi and Oden's model can further be simplified by noticing that the continuousness of the purely-normal stress at the boundary; however, from our numerical analysis, we find that this reduced model is non-convergent in a finite-difference setting. Thus, we insist upon the given formulation. 

\subsection{A Numerical Example}

To proceed with our analysis, we numerically model the overlying body as a three-dimensional body and we do not approximate this body as a shell or otherwise. Thus, the displacement at the contact region with this approach is obtained by the use of the standard equilibrium equations in linear elasticity and the modified Kikuchi and Oden's model.\\

In accordance with the framework that is introduced in section \ref{ch4Num}, the overlying body is restricted to the region $x^3 \in (0,h)$. Now, we can express the governing equations of the overlying body as follows
\begin{align*}
(\lambda + \mu)\partial_2(\bar \nabla_{\!i} v^i ) + \mu (\bar\psi_2)^2 \bar \Delta v^2 & = 0,\\
(\lambda + \mu)\partial_3(\bar \nabla_{\!i} v^i ) + \mu \bar \Delta v^3 & = 0,
\end{align*}
where $\boldsymbol v = \boldsymbol(0,v^2(x^2,x^3),v^3(x^2,x^3) \boldsymbol )$ is the displacement field of the overlying body, the perturbed governing equations of the overlying body as follows
\begin{align*}
(\lambda + \mu)\partial_2(\bar \nabla_{\!i} \delta v^i) + \mu (\bar\psi_2)^2 \bar \Delta  \delta v^2 & = 0,\\
( \lambda +  \mu)\partial_3(\bar \nabla_{\!i} \delta v^i) + \mu \bar \Delta \delta v^3 & = 0,
\end{align*}
where $\delta\boldsymbol v = \boldsymbol (0,\delta v^2(x^2,x^3),\delta v^3(x^2,x^3)\boldsymbol )$ is a small perturbation of the displacement field of the overlying body, and the perturbed governing equations of the foundation as follows
\begin{align*}
(\bar\lambda + \bar\mu)\partial_2(\bar \nabla_{\!i} \delta u^i ) + \bar \mu (\bar\psi_2)^2 \bar \Delta  \delta u^2 & = 0,\\
(\bar \lambda + \bar \mu)\partial_3(\bar \nabla_{\!i} \delta u^i ) + \bar \mu \bar \Delta \delta u^3 & = 0,
\end{align*}
where $\delta\boldsymbol u = \boldsymbol (0,\delta u^2(x^2,x^3),\delta u^3(x^2,x^3)\boldsymbol )$ is the perturbation of the displacement field of the foundation. Also, we can express the boundary conditions of the overlying body as follows
\begin{align*}
\big[(\lambda + 2\mu) \partial_2 v^2 + \lambda\left( \partial_3 v^3 + \bar\Gamma^2_{\!22} v^2 + \bar\Gamma^2_{\!23} v^3\right)\big]|_{\{\partial \omega_{T_0}^\text{New}, \partial \omega_{T_\text{max}}^\text{New}\}\times [0,h]} 
&= \{\tau_0 , \tau_\text{max}\} ,  \\
\big[\lambda\left( \partial_2 v^2 + \bar\Gamma^2_{\!22} v^2 + \bar\Gamma^2_{\!23} v^3\right) + (\lambda + 2\mu) \partial_3 v^3\big]|_{ (-\frac{1}{2}\pi,\frac{1}{2}\pi)\times\{h\}}
& = 0, \\
\big[(\bar\psi_2)^2\partial_3v^2 +\partial_2v^3\big] |_{\{\partial \omega^\text{New}\times [0,h]\}\cup \{ (-\frac{1}{2}\pi,\frac{1}{2}\pi)\times\{h\}\}} 
&= 0 ,
\end{align*}
boundary conditions of the displacement fields  as follows
\begin{align*}
\big[v^3-u^3\big]|_{ \omega^\text{New}} & =0 ~\text{(continuous radial displacement)},\\
\big[T^3_3(\boldsymbol v) -T^3_3(\boldsymbol u)\big] |_{\omega^\text{New}}&= 0~\text{(continuous radial stress)},
\end{align*}
and the boundary conditions of the perturbations as follows,
\begin{align*}
\delta u^2|_{ \partial \Omega_0^\text{New} \cup \overline{\partial \Omega}_f^\text{New}} &= 0,\\
\delta u^3|_{\overline{\partial \Omega}^\text{New}} &= 0,\\
\delta v^2|_{\{\partial \omega^\text{New}\times (0,h)\}\cup \{ [-\frac{1}{2}\pi,\frac{1}{2}\pi]\times\{h\}\}}&= 0,\\
\delta v^3 |_{\bar\omega^\text{New} \cup\{\partial \omega^\text{New}\times (0,h)\}\cup \{ [-\frac{1}{2}\pi,\frac{1}{2}\pi]\times\{h\}\}} &= 0.
\end{align*}
Thus, the equations characterising the frictional coupling of the overlying body to the foundation can be expressed as follows:\\
If $\psi_2|v^2 - u^2||_{\omega^\text{New}}\geq\epsilon$, then
\begin{align*}
\big[\mu \left(\psi_2\partial_3v^2 + (\psi_2)^{-1} \partial_2 v^3  \right) + \nu_F \mathrm{sign}(v^2 - u^2)T^3_3(\boldsymbol v)  \big]|_{\omega^\text{New}} &= 0,\\
\big[\bar\mu \left(\psi_2\partial_3u^2 + (\psi_2)^{-1} \partial_2 u^3  \right) + \nu_F \mathrm{sign}(v^2 - u^2)T^3_3(\boldsymbol u)  \big]|_{\omega^\text{New}} &= 0;
\end{align*}
If $\psi_2|v^2- u^2||_{\omega^\text{New}}<\epsilon$, then
\begin{align*}
\big[\mu \left(\psi_2\partial_3\delta v^2  \right) +\nu_F \epsilon^{-1} \psi_2(v^2 - u^2)T^3_3(\delta \boldsymbol v) + \nu_F \epsilon^{-1}\psi_2(\delta v^2-\delta u^2) T^3_3(\boldsymbol v) &\\
+ \mu \left(\psi_2\partial_3v^2 + (\psi_2)^{-1} \partial_2 v^3  \right) + \nu_F \epsilon^{-1}\psi_2( v^2 - u^2) T^3_3(\boldsymbol v) &\big] |_{\omega^\text{New}}= 0,\\
\big[\bar\mu \left(\psi_2\partial_3\delta u^2  \right) +\nu_F \epsilon^{-1}\psi_2 (v^2-u^2)T^3_3(\delta \boldsymbol u)+ \nu_F \epsilon^{-1}\psi_2(\delta v^2-\delta u^2) T^3_3(\boldsymbol u) &\\
+ \bar\mu \left(\psi_2\partial_3u^2 + (\psi_2)^{-1} \partial_2 u^3  \right) + \nu_F \epsilon^{-1}\psi_2( v^2-u^2) T^3_3(\boldsymbol u) & \big] |_{\omega^\text{New}}= 0,
\end{align*}
where
\begin{align*}
T^3_3(\boldsymbol v) & =\lambda\left( \partial_2 v^2 + \bar\Gamma^2_{\!22} v^2 + \bar\Gamma^2_{\!23} v^3 \right) + (\lambda + 2\mu) \partial_3 v^3,\\
T^3_3 (\boldsymbol u)& =\bar\lambda\left( \partial_2 u^2 + \bar\Gamma^2_{\!22} u^2 + \bar\Gamma^2_{\!23} u^3 \right) + (\bar\lambda + 2\bar\mu) \partial_3 u^3.
\end{align*}

To conduct numerical experiments, we use the second-order accurate iterative-Jacobi finite-difference method with Newton's method for nonlinear systems (see chapter 10 of  Burden \emph{et al.} \cite{burden2015numerical}). Also, as a result of the grid dependence in the overlying body, we must satisfy the condition $ \psi_0 \Delta x^2 \leq \Delta x^3$, $\forall ~ \psi_0 \in \{\bar\psi_2(x^2,x^3) \mid x^2\in[-\frac{1}{2}\pi,\frac{1}{2}\pi] ~\text{and}~ x^3\in[0,h] \}$. For our purposes, we let $\Delta x^2 = \frac{1}{N-1}\pi$ and $\psi_0 = \bar\psi_2(\frac{1}{4}\pi,h)$, where $N=250$.\\

\begin{figure*}[!h]
\centering
\includegraphics[trim = 2cm 1cm 2cm 1cm, clip = true, width=1\linewidth]{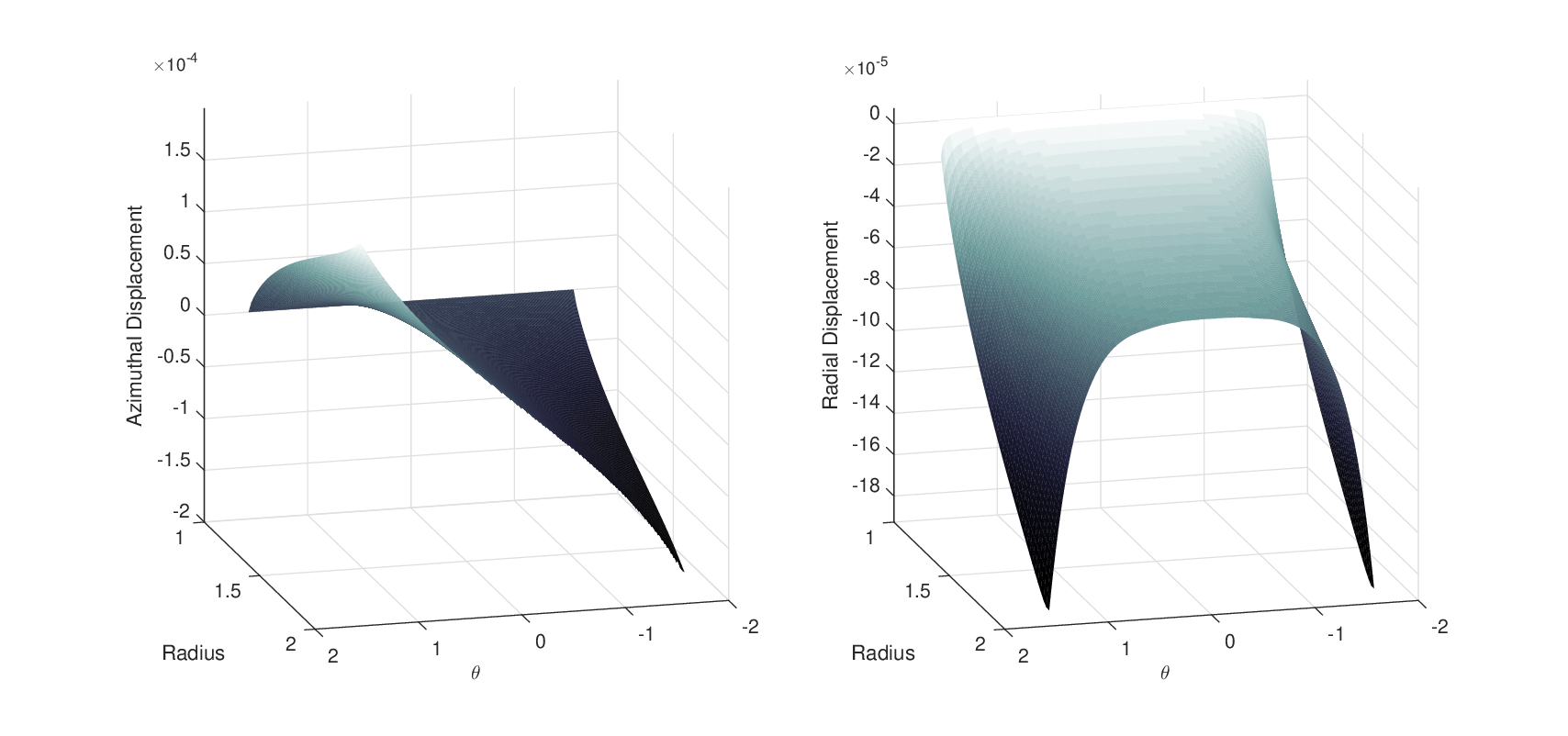}
\caption{Displacement field of the foundation of the modified Kikuchi and Oden's model, where $\theta = x^2$\label{Ch4Oden}}
\end{figure*}

Fig. \ref{Ch4Oden} is calculated with the values $\tau_\text{max}= 1$N/$\text{m}^2$, $b=2$m, $h=\frac{1}{8}$m, $ E = 8000$Pa, $\nu = \frac{1}{4}$, $\varepsilon = 10^{-10}$m, and $\nu_F = 1$, and it shows the azimuthal (i.e. $u^2$) and the radial (i.e. $u^3$) displacements of the foundation. The maximum azimuthal displacements are observed at $x^2=\pm\frac{1}{2}\pi$ with respective azimuthal displacements of $u^2 =\pm 1.70\times 10^{-4}$rad. The maximum radial displacement is observed at $x^2=\pm\frac{1}{2}\pi$ with a radial displacement of $u^3 =-1.85\times 10^{-4}$m. Also, in the interval $x^2\in (- \frac12\pi,\frac12\pi)$, i.e. in the entire contact region, we see that the overlying body is at limiting-equilibrium. Just as it is in the analysis of Fig. \ref{Ch4Shell}, these observations simply imply that the shell is more likely to debond from foundation at the boundaries where we apply external stress, and more likely to stay bonded away from those boundaries. However, Fig. \ref{Ch4Oden} predicts a higher likelihood of debonding relative to the shell model as now the entire contact region is at limiting-equilibrium for the two-body Coulomb's law of static friction.

\subsection{Comparing a Shell Frictionally Coupled to an Elastic Foundation and Two-Body Coulomb's Law of Static Friction}

Our final goal in this section is to investigate how our model for a shell on an elastic foundation with friction predicts the displacement field of the foundation relative to the two-body elastic model with friction. From this, we should be able to ascertain how the stresses from the thin body (approximated by a shell or otherwise) propagate to the foundation and deforms it, and we do this for the variables  $\nu_F$, $\delta \tau =\tau_\text{max}/\tau_0$, $\delta b = b/a$, $\delta h = h/L$, $\delta E =E/\bar E$ and $\delta \nu = \nu/\bar \nu$. To calculate the relative error between the displacement field of the foundation predicted by each model, we define the following metric
\begin{align*}
\mathrm{RelativeError}(u^i) = \frac{\sqrt{\sum_{\{k,l\}}||u_\text{shell-model}^i(y^2_k, y^3_l)-u_\text{Kikuchi-Oden}^i(y^2_k, y^3_l)||^2}}{\sqrt{\sum_{\{k,l\}}||u_\text{shell-model}^i(y^2_k, y^3_l)||^2+||u_\text{Kikuchi-Oden}^i(y^2_k, y^3_l)||^2}},
\end{align*}
where $y^2_k = -\frac12\pi +k\Delta x^2$, $y^3_l = -L+l\Delta x^3$, $0\leq k \leq (N-1)$ and   $0\leq l \leq (N-1)$. Note that we assume the default values  $\nu_F=1$, $\delta \tau =1 $, $\delta b=1$, $\delta h =\frac18$, $\delta E = 8$ and $ \delta \nu = 1$ and $\varepsilon = 10^{-10}$m throughout, unless it strictly says otherwise.\\

\begin{figure}[!h]
\centering
\includegraphics[trim = 2cm 1cm 2cm 1cm , clip = true, width=1\linewidth]{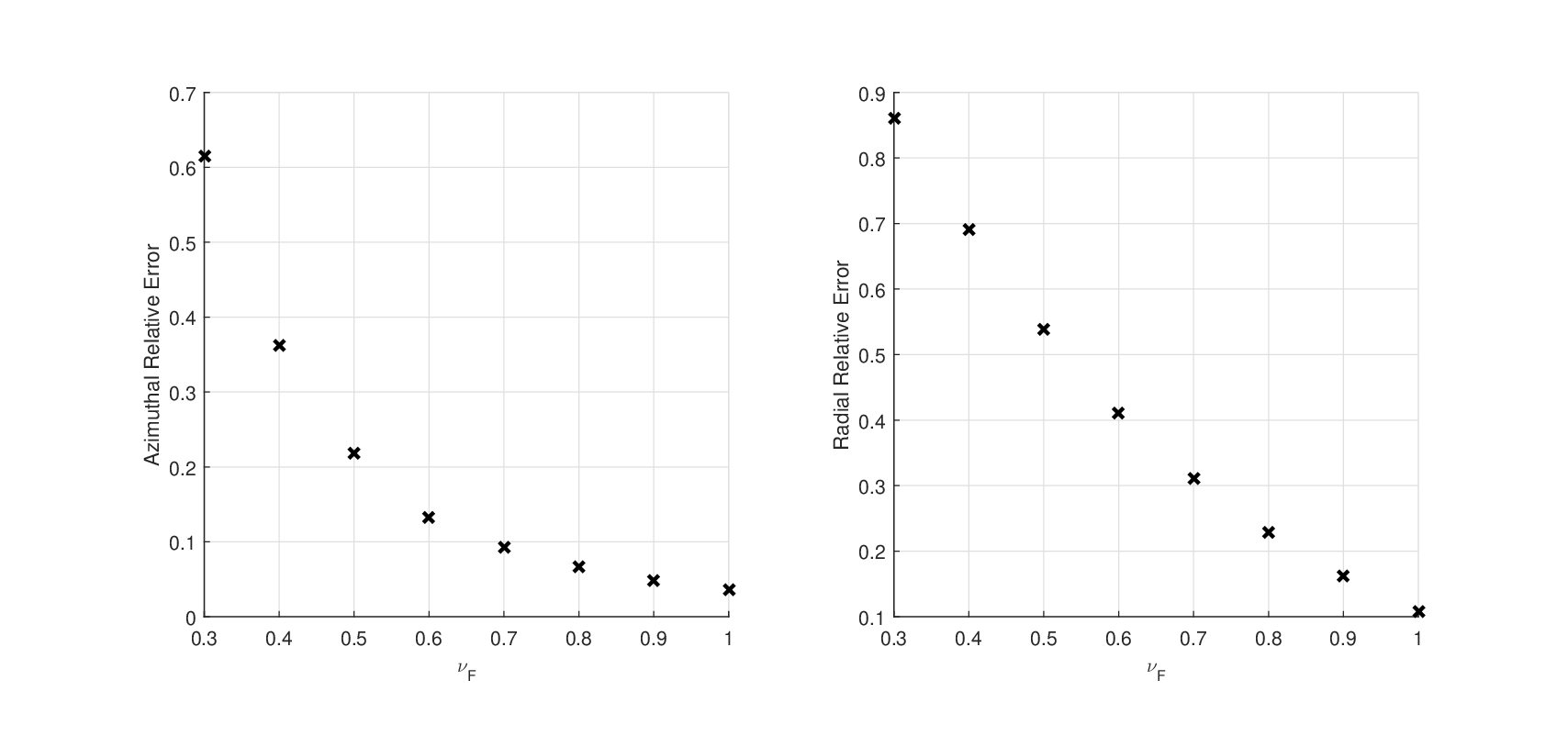}
\caption{Relative error for $\nu_F $\label{Ch4mu}}
\end{figure}

\begin{figure}[!h]
\centering
\includegraphics[trim = 2cm 1cm 2cm 1cm, clip = true, width=1\linewidth]{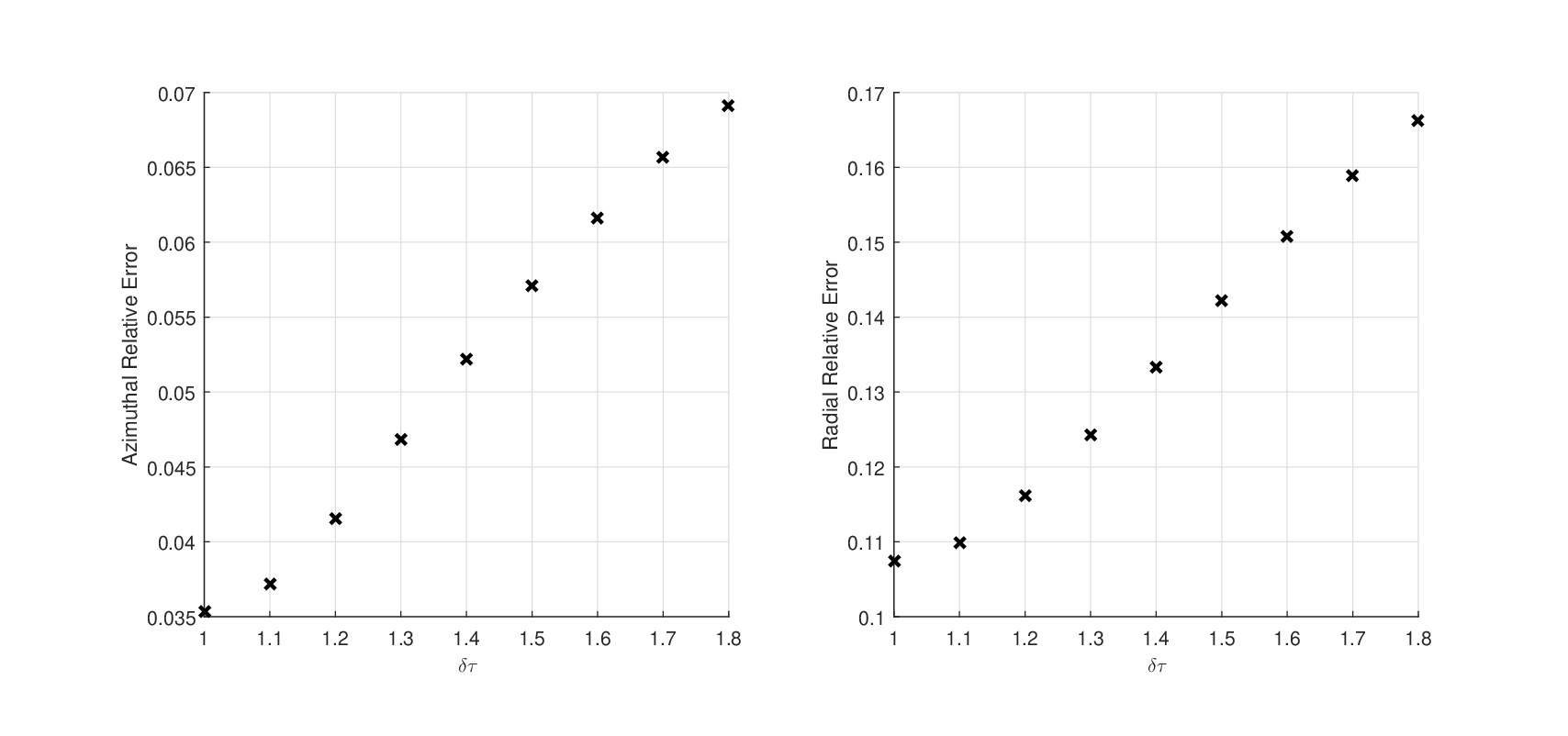}
\caption{Relative error for $\delta \tau$\label{Ch4T}}
\end{figure}

\begin{figure}[!h]
\centering
\includegraphics[trim = 2cm 1cm 2cm 1cm, clip = true, width=1\linewidth]{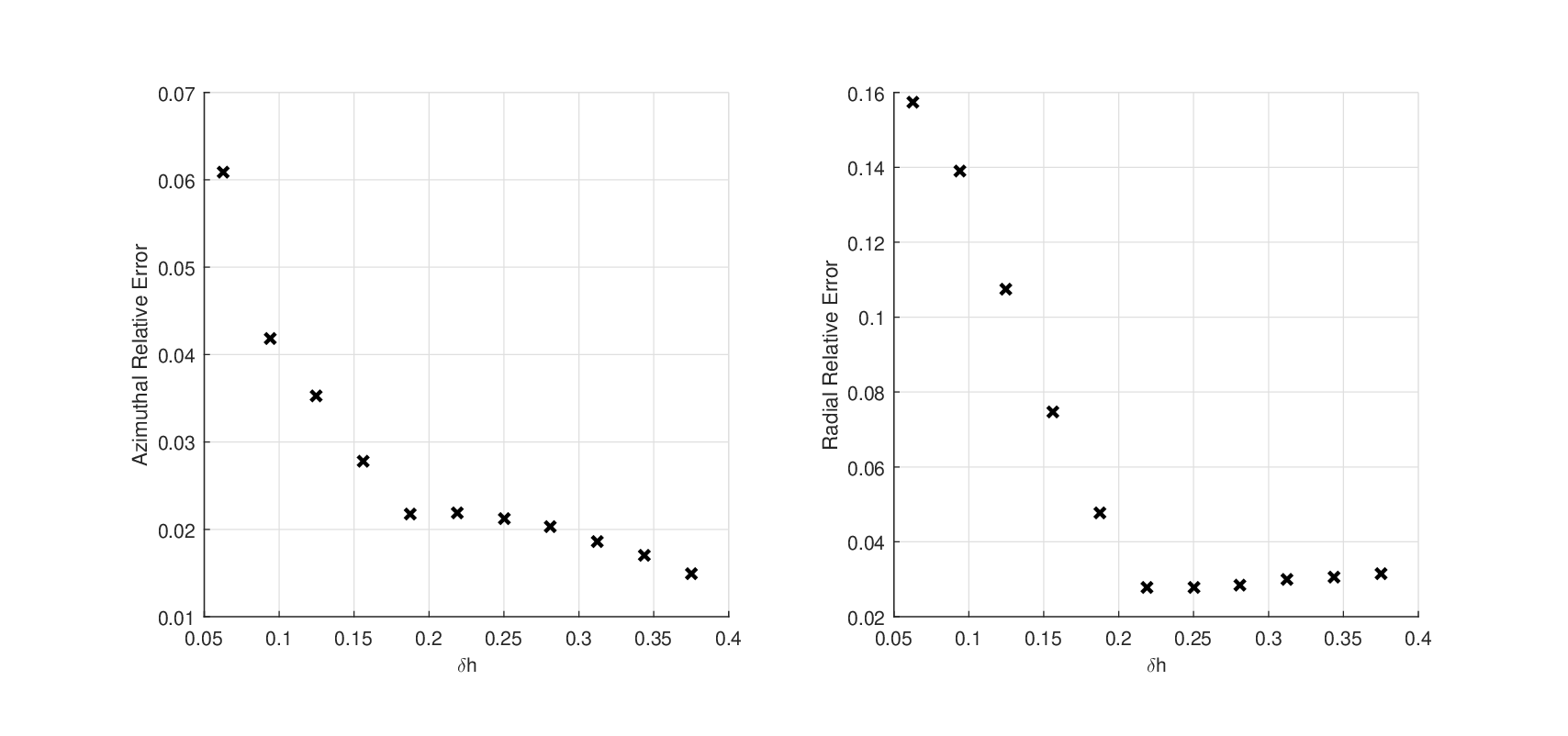}
\caption{Relative error for $\delta h$\label{Ch4h}}
\end{figure}

\begin{figure}[!h]
\centering
\includegraphics[trim = 2cm 1cm 2cm 1cm, clip = true, width=1\linewidth]{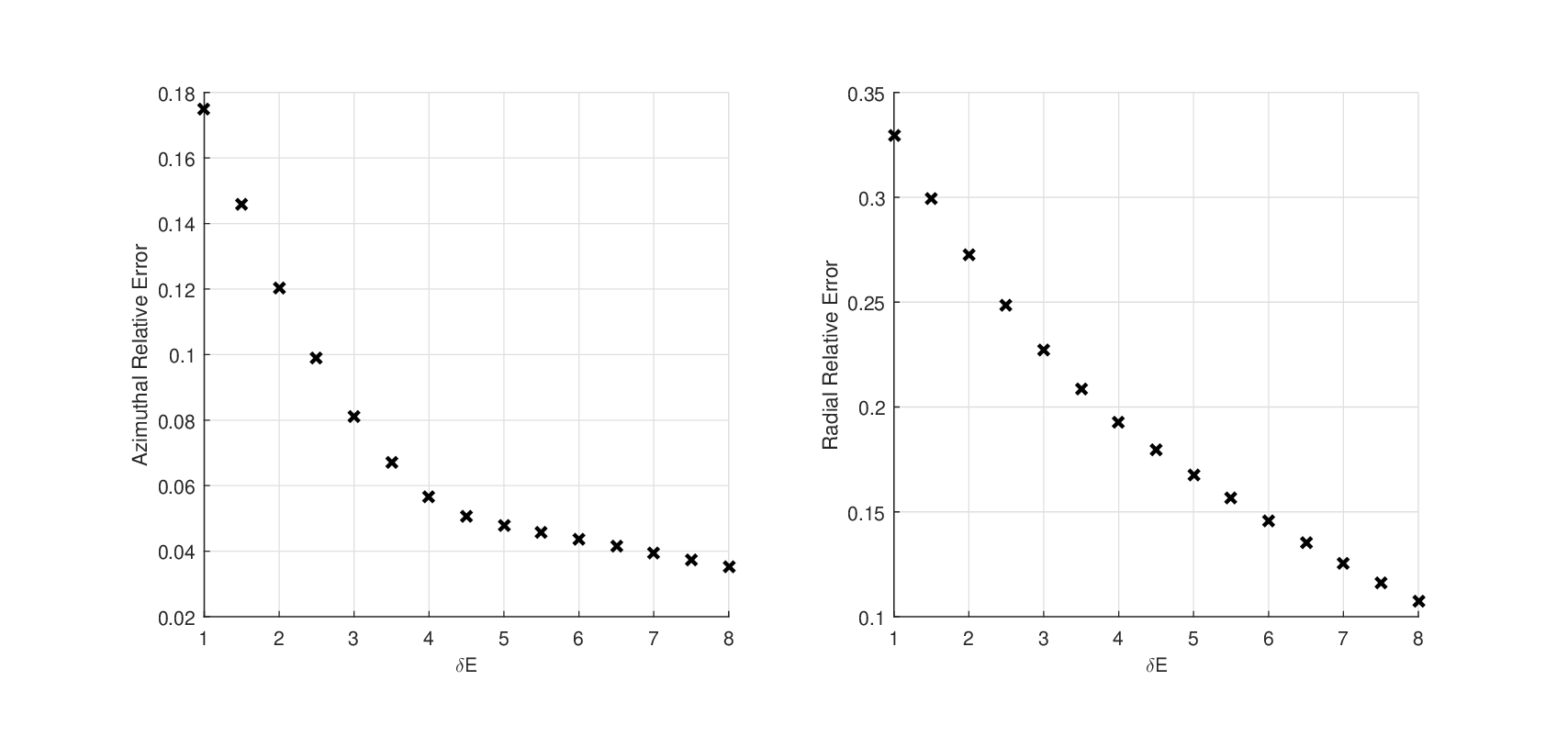}
\caption{Relative error for $\delta E$\label{Ch4E}}
\end{figure}

\begin{figure}[!h]
\centering
\includegraphics[trim = 2cm 1cm 2cm 1cm, clip = true, width=1\linewidth]{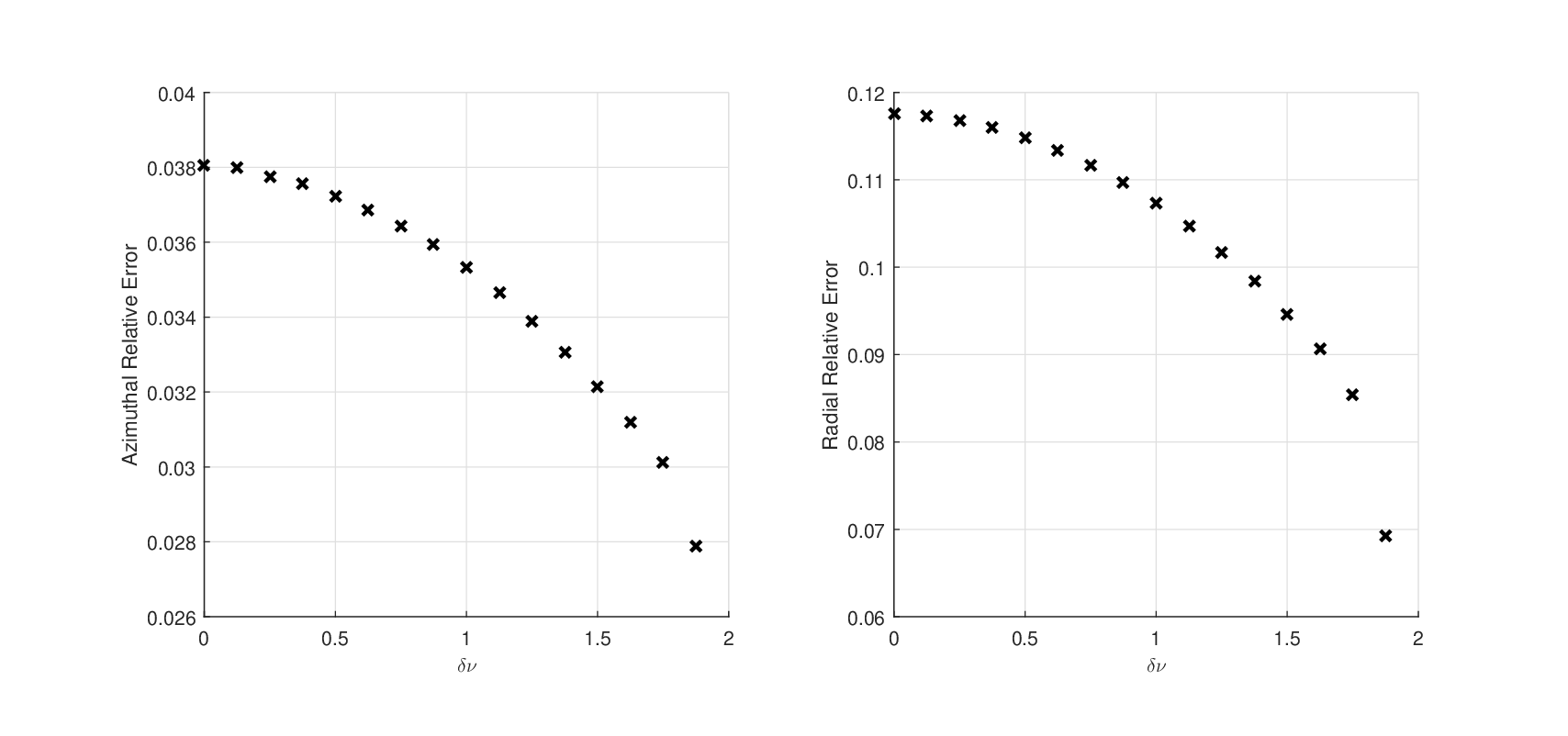}
\caption{Relative error for $\delta \nu$\label{Ch4nu}}
\end{figure}

\begin{figure}[!h]
\centering
\includegraphics[trim = 2cm 1cm 2cm 1cm, clip = true, width=1\linewidth]{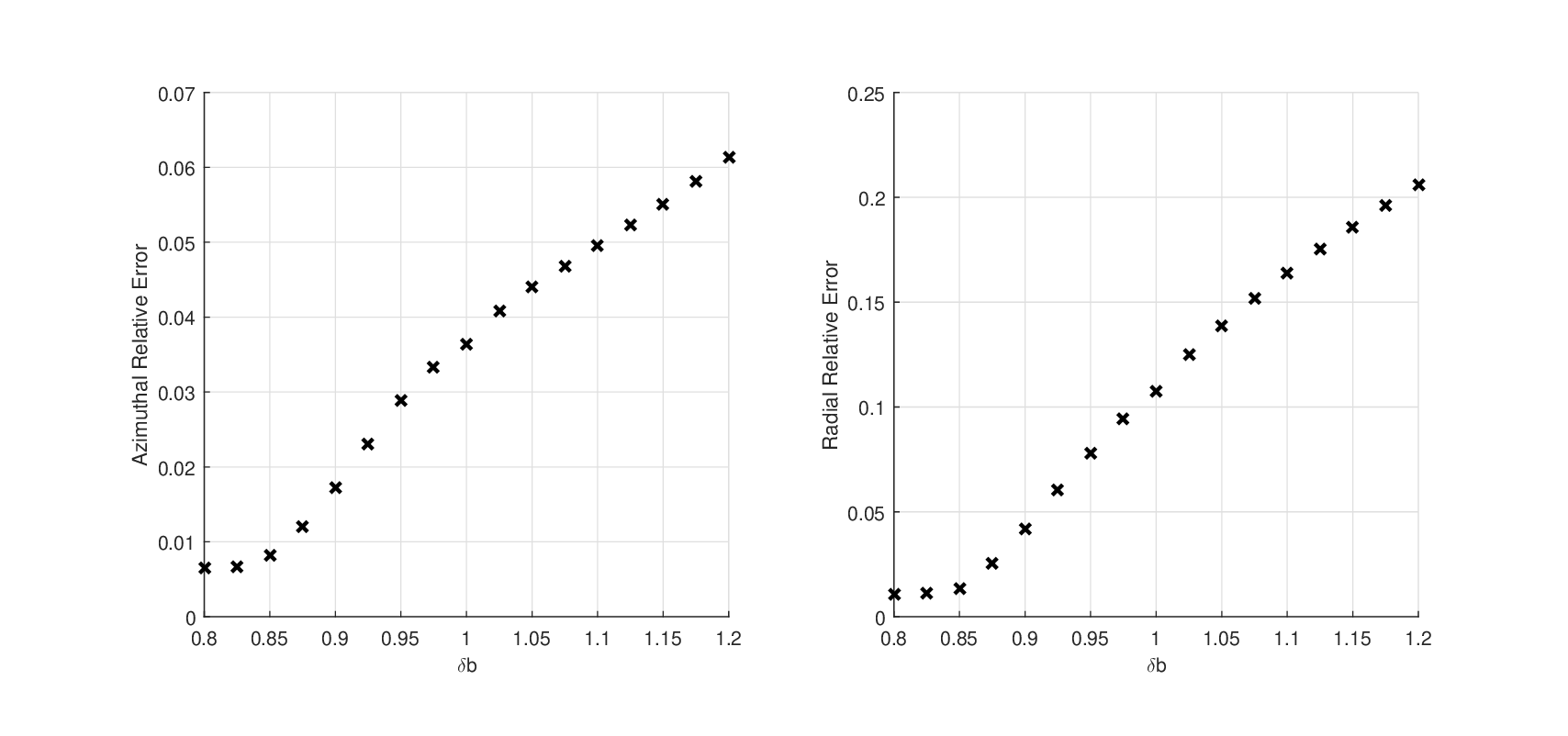}
\caption{Relative error for $\delta b$\label{Ch4c}} 
\end{figure}

Fig. \ref{Ch4mu} shows that as the coefficient of friction at the contact region, $\nu_F$, increases, the relative errors decrease, and this reduction in the error is significant in its magnitude. This implies that rougher the contact surface is, then closer our shell model with friction resembles the modified Kikuchi and Oden's model. This is an intuitive result as the coefficient of friction increases, both models resemble the bonded case, and in chapter 3 of Jayawardana \cite{jayawardana2016mathematical}, it is shown that the our bonded shell model is a better approximation of the overlying body with respect to the asymptotic model implied by Baldelli and Bourdin's method \cite{Andres}. \\

Fig. \ref{Ch4T} shows that as the traction ratio, $\delta \tau$, increases, both azimuthal and radial relative errors also increase. This implies that the limiting-equilibriums implied by each model can be very different. We observed this effect in our earlier numerical modelling (recall the analysis of Fig. \ref{Ch4Shell} and Fig. \ref{Ch4Oden}).\\

Fig. \ref{Ch4h} shows that as the relative thickness of the shell, $\delta h$, decreases, both the azimuthal and the radial relative errors increase. This is contradicts the derivation assumptions of our shell model with friction as we derived our displacement-based friction condition by considering Coulomb's law to remain valid in the limit $\delta h \to 0$, and thus, we should expect a better agreement between the two models for smaller values of $\delta h$. \\

Fig. \ref{Ch4E} shows the that as the relative Young's modulus of the shell, $\delta E$, increases, the relative errors decrease. If one assumes the shell is bonded to the elastic foundation, then this result seems to be consistent with Aghalovyan's asymptotic analysis of the modulus of an orthotropic foundation and two-layer anisotropic plates \cite{aghalovyan2015asymptotic}.\\

Fig. \ref{Ch4nu} shows that as the relative Poisson's ratio of the shell, $\delta \nu$, increases, the relative error decreases. This implies that as the shell becomes incompressible, both models would be in better agreement. \\

Fig. \ref{Ch4c} shows that as $\delta b$ decreases, both azimuthal and radial relative errors also decrease. Assuming the contact region $x^2\in(-\frac{1}{2}\pi,\frac{1}{2}\pi)$ (with $x^2\neq\pm\frac{1}{2}\pi)$, this result may be interpreted as follows: as the radius of curvature of the contact region increases, the relative errors decrease. Note that we derived our shell equation to be valid for contact regions with high radius of curvatures (see condition \ref{dfnShell}).

\begin{remark}
In a different study, to ascertain the physical validity of our shell model with friction, we conduct human trials to measure the frictional interactions that fabrics have on soft-tissue of human subjects: 10 subjects in the first trial and 8 subjects in the second trial (see chapter 6 of Jayawardana \cite{jayawardana2016mathematical}). We discover the following: (i) a positive correlation between the displacement of soft-tissue and the volume of soft-tissue; (ii) a positive correlation between the applied tension to the fabric and the volume of soft-tissue; (iii) a negative correlation between the displacement of the soft-tissue and the Young's modulus of soft-tissue; and (iv) a negative correlation between  the applied tension to the fabric and the Young's modulus of soft-tissue. From further numerical modellings of our shell model with friction (where now the shell is approximated with a shell-membrane) we were able to predict the correlations-(i) to (iii). However, our numerical modelling could not predict correlation-(iv). It is unclear whether the discrepancy in correlation-(iv) is due to a flaw in our numerical modelling (see section 3 of Jayawardana \emph{et al.} \cite{jayawardana2017quantifying}) or our data analysis of the experiments.
\end{remark}

\section{Conclusions}

In our analysis, we derived a model for a shell that is frictionally coupled to an elastic foundation. We used Kikuchi and Oden's  model  for Coulomb's law of static friction \citep{Kikuchi} to derive a displacement-based static-friction condition. By construction, this displacement-based friction condition is mathematically sound as we proved the existence and the uniqueness of solutions for our shell model with friction with the aid of the work of Kinderlehrer and Stampacchia \cite{kinderlehrer2000introduction} (and section 8.4.2 of Evans \cite{Evans}). Note that, as far as we are aware, this is the first derivation of a displacement-based friction condition, as only force, stress and energy based friction conditions currently exist in the literature. \\

For numerical analysis, we modified Kikuchi and Oden's model for Coulomb's law of static friction \cite{Kikuchi} to model a full two-body elasticity contact problem in curvilinear coordinates. The purpose of numerical analysis is to ascertain how the displacement field of the foundation behaves when the overlying body is modelled with our shell model with friction in comparison to when the overlying body is modelled with standard equilibrium equations in linear elasticity and modified Kikuchi and Oden's model. Note that, as far as we are aware, this is the first derivation of a two-body 3D (and 2D) elasticity contact problem with friction, as only one-body elasticity contact problems (one elastic-body in contact with one rigid-body) and two-body 1D elasticity contact problems  (see analysis of elastic strings in contact  \cite{doonmez2004model,grandgeorge2021mechanics,maddocks1987ropes,warren2018clothes}) with friction, currently exist in the literature. The numerical results indicate that, if the shell has a relatively high Young's modulus (i.e. stiff) or has a relatively high Poisson's ratio (i.e. close to incompressible), and the contact region has a high coefficient of friction or has a high radius of curvature, then the displacement field of the foundation predicted by both models are in better agreement. We also observed that both models are in better agreement for thicker shells, which is a contradictory result as it is inconsistent with the derivation of our shell model with friction; therefore, further research is needed to resolve this contradiction.\\

From our numerical analysis, the greatest reduction in the error is observed for higher coefficients of friction, i.e. as the coefficient of friction increases, the bodies behaves as if they are bonded, and thus, the greater agreement between the solutions of both models. This is an expected result as our overlying shell model was initially derived to approximate bonded thin bodies on elastic foundations, and the efficacy of this model is numerically demonstrated in sections 3.5 and 3.6 of  Jayawardana \cite{jayawardana2016mathematical}.  The second greatest reduction in the error is observed for higher Young's moduli of the shell. This is also an expected result as it seems to be consistent with the  asymptotic analysis of similar problems (given that the shell is bonded to the elastic foundation) by Aghalovyan \cite{aghalovyan2015asymptotic}.\\

On a final note, given that the thin body is further approximated by a shell-membrane (i.e. neglecting bending effects of the shell), our model can be use to investigate the frictional interactions between fabrics and human soft-tissue (i.e. the stresses on human skin and the deformation of the subcutaneous tissue due to friction generated by everyday attire). A detail study of this can be found in chapter 6 of Jayawardana \cite{jayawardana2016mathematical} and Jayawardana \emph{et al.} \cite{jayawardana2017quantifying}.\\

\section*{Acknowledgments}
We thank Dr Nick Ovenden (UCL) and Prof Alan Cottenden (UCL) for their supervision, Brad Turner (TEKOR) for his assistance, and Christopher Law for the illustrations.

\bibliographystyle{./model1-num-names}
\bibliography{ShellsOnElasticFoundationsWithFriction}%

\begin{thebibliography}{10}
\expandafter\ifx\csname url\endcsname\relax
  \def\url#1{\texttt{#1}}\fi
\expandafter\ifx\csname urlprefix\endcsname\relax\def\urlprefix{URL }\fi
\expandafter\ifx\csname href\endcsname\relax
  \def\href#1#2{#2} \def\path#1{#1}\fi

\bibitem{Kikuchi}
N.~Kikuchi, J.~T. Oden, Contact problems in elasticity: a study of variational
  inequalities and finite element methods, SIAM, 1988.

\bibitem{kinderlehrer2000introduction}
D.~Kinderlehrer, G.~Stampacchia, An introduction to variational inequalities
  and their applications, SIAM, 2000.

\bibitem{jayawardana2016mathematical}
K.~Jayawardana, Mathematical theory of shells on elastic foundations: an
  analysis of boundary forms, constraints, and applications to friction and
  skin abrasion, Ph.D. thesis, UCL (University College London) (2016).

\bibitem{bowden2001friction}
F.~P. Bowden, F.~P. Bowden, D.~Tabor, The friction and lubrication of solids,
  Vol.~1, Oxford university press, 2001.

\bibitem{clark1981mechanics}
S.~K. Clark, Mechanics of pneumatic tires, US Government Printing Office, 1981.

\bibitem{jayawardana2017quantifying}
K.~Jayawardana, N.~C. Ovenden, A.~Cottenden, Quantifying the frictional forces
  between skin and nonwoven fabrics, Frontiers in physiology 8 (2017) 107.

\bibitem{bergfeld1985trauma}
W.~F. Bergfeld, J.~S. Taylor, Trauma, sports, and the skin, American journal of
  industrial medicine 8~(4-5) (1985) 403--413.

\bibitem{asserin2000measurement}
J.~Asserin, H.~Zahouani, P.~Humbert, V.~Couturaud, D.~Mougin, Measurement of
  the friction coefficient of the human skin in vivo: quantification of the
  cutaneous smoothness, Colloids and surfaces B: Biointerfaces 19~(1) (2000)
  1--12.

\bibitem{levit1977jogger}
F.~Levit, Jogger's nipples., The New England journal of medicine 297~(20)
  (1977) 1127--1127.

\bibitem{wilkinson1985dermatitis}
D.~S. Wilkinson, Dermatitis from repeated trauma to the skin, American journal
  of industrial medicine 8~(4-5) (1985) 307--317.

\bibitem{maklebust2001pressure}
J.~Maklebust, M.~Sieggreen, Pressure ulcers: Guidelines for prevention and
  management, Lippincott Williams \& Wilkins, 2001.

\bibitem{shrank1979aetiology}
A.~B. Shrank, The aetiology of juvenile plantar dermatosis, British Journal of
  Dermatology 100~(6) (1979) 641--648.

\bibitem{johnson1987contact}
K.~L. Johnson, Contact mechanics, Cambridge university press, 1987.

\bibitem{quadling2004mechanics}
D.~Quadling, H.~Neill, Mechanics 1, Cambridge Advanced Level Mathematics for
  OCR, Cambridge University Press, 2004.

\bibitem{badiale2010semilinear}
M.~Badiale, E.~Serra, Semilinear Elliptic Equations for Beginners: Existence
  Results via the Variational Approach, Springer Science \& Business Media,
  2010.

\bibitem{ciarlet2005introduction}
P.~G. Ciarlet, An introduction to differential geometry with applications to
  elasticity, Journal of Elasticity 78~(1) (2005) 1--215.

\bibitem{rao2003engineering}
C.~L. RAO, J.~Lakshinarashiman, R.~Sethuraman, S.~M. Sivakumar, Engineering
  Mechanics: Statics and Dynamics, PHI Learning Pvt. Ltd., 2003.

\bibitem{gao2000finite}
D.~Y. Gao, Finite deformation beam models and triality theory in dynamical
  post-buckling analysis, International journal of non-linear mechanics 35~(1)
  (2000) 103--131.

\bibitem{doonmez2004model}
S.~D{\"o}onmez, A.~Marmarali, A model for predicting a yarn's knittability,
  Textile research journal 74~(12) (2004) 1049--1054.

\bibitem{grandgeorge2021mechanics}
P.~Grandgeorge, C.~Baek, H.~Singh, P.~Johanns, T.~G. Sano, A.~Flynn, J.~H.
  Maddocks, P.~M. Reis, Mechanics of two filaments in tight orthogonal contact,
  Proceedings of the National Academy of Sciences 118~(15) (2021).

\bibitem{maddocks1987ropes}
J.~H. Maddocks, J.~B. Keller, Ropes in equilibrium, SIAM Journal on Applied
  Mathematics 47~(6) (1987) 1185--1200.

\bibitem{warren2018clothes}
P.~B. Warren, R.~C. Ball, R.~E. Goldstein, Why clothes don’t fall apart:
  Tension transmission in staple yarns, Physical review letters 120~(15) (2018)
  158001.

\bibitem{adams2003sobolev}
R.~A. Adams, J.~J. Fournier, Sobolev spaces, Elsevier, 2003.

\bibitem{kay1988schaum}
D.~C. Kay, Schaum's Outline of Tensor Calculus, McGraw Hill Professional, 1988.

\bibitem{Evans}
L.~C. Evans, Partial differential equations, Graduate studies in mathematics
  19~(4) (1998) 7.

\bibitem{schilling2017measures}
R.~L. Schilling, Measures, integrals and martingales, Cambridge University
  Press, 2017.

\bibitem{reddy2006theory}
J.~N. Reddy, Theory and analysis of elastic plates and shells, CRC press, 2006.

\bibitem{moon2012field}
P.~Moon, D.~E. Spencer, Field theory handbook: including coordinate systems,
  differential equations and their solutions, Springer, 2012.

\bibitem{burden2015numerical}
R.~Burden, J.~Faires, A.~Burden, Numerical Analysis, Cengage Learning, 2015.

\bibitem{Andres}
A.~A.~L. Baldelli, B.~Bourdin, On the asymptotic derivation of winkler-type
  energies from 3d elasticity, Journal of Elasticity 121~(2) (2015) 275--301.

\bibitem{aghalovyan2015asymptotic}
L.~A. Aghalovyan, Asymptotic theory of anisotropic plates and shells, World
  Scientific, 2015.

\end{thebibliography}
\biboptions{sort&compress}


\end{document}